Resubmision to Physical Review D

Manuscript code: DT6378
Originally from: aneronov@mech.math.msu.su

Figure updates: PostScript figures included in this electronic submission

Details of changes:
1. Number of figures is reduced from 15 to 8, as it was required by
referee. 

2. Fractions within fractions in formulas are removed almost everywhere
in accordance with referees requirement, except for the expressions on 
pages 29, 30 and 31 which describe the spectrum. It is because in this 
unchanged (although not very nice-looking) form they are easily identified
with expressions for spectrum obtained in Phys.Rev.D57,1118(1998) and I'd
like it to be clear.

3. The text has minor corrections in order to correspond to the reduced 
number of figures.

Response to recommendations and criticisms:
Thank you very much for your kind letter and good assessment to my work.
The remarks that you made consern the form and not the essence of 
the paper, so 
I have corrected the text and reduced number of figures 
following the recommendations 
made by referee.

Remarks intended solely for the editor:

\documentstyle[preprint,aps]{revtex}

\begin{document}
\draft
\preprint{}
\bibliographystyle{prsty.bst}
 \title{ Quasiclassical mass spectrum of the black hole model with
selfgravitating dust shell.}
 \author{A.~Yu.~Neronov}
 \address{
Department of Mathematics and Mechanics, Lomonosov Moscow State
University, 119899, Moscow, Russia\\ e-mail:
aneronov@mech.math.msu.su }
 \date{\today}
 \maketitle
 \begin{abstract}
We consider a quantum mechanical black hole model introduced in
{\it Phys.Rev.}, {\bf D57}, 1118 (1998) that consists of the
selfgravitating dust shell. The Schroedinger equation for
this model is a finite difference equation with the shift of the argument
along the imaginary axis. Solving this equation in quasiclassical limit in
complex domain leads to quantization conditions that define discrete
quasiclassical mass spectrum. One of the quantization conditions is
Bohr-Sommerfeld condition for the  bound  motion of the shell. The other
comes from the requirement that the wave function is unambiguously defined
on the Riemannian surface on which the coefficients of Schroedinger
equation are regular. The second quantization condition remains valid for
the unbound motion of the shell as well, and in the case of a collapsing
null-dust shell leads to $m\sim\sqrt{k}$ spectrum.
 \end{abstract}

 \narrowtext
 \section*{Introduction.}
The black hole
physics gives us an example of the strong gravitational fields. The existence
of the event (apparent) horizons causes the Hawking's evaporation of the black
 holes . The fate of the evaporating black holes becomes a subject of
interest. The quantum theory may throw some light on many problems of the
classical black hole physics. In the absence of the
theory of quantum gravity we have to construct
a new theory every time we want to quantize some classical gravitating
object.

The simplest black
holes models which nevertheless capture some important properties of the
system consisting of the gravitational field interacting with matter are
of special importance in the theory of black holes. Many authors have
considered the models where the interaction of the gravitational field
with matter in the form of a thin spherically symmetric shell was taken in
account in the selfconsistent way \cite{shells,shell,berezin,louko}.

In \cite{prd57} the classical geometrodynamics for the system of a thin
dust matter shell in its own gravitational field was constructed. The
canonical transformation was found which reduce the system of constraints
to a rather simple system of functions on the phase space. The constraints
are easily realized on the quantum level. Quantization of such a model
in the coordinate representation leads to the Schroedinger equation in
finite differences. The shift in the
argument is along imaginary axis which has very important consequences.
One of them is that the wave functions which are the solutions to such an
equation should be analytical functions on the appropriate Riemannian
surface.

In the ordinary quantum mechanics we are dealing with the second order
differential equations. We demand that the solution should be at least
two times differentiable. To find eigenfunctions and spectrum we need
to specify a class of functions, usually by imposing appropriate boundary
conditions.  In the  case of the  finite differences operator with the shift
along the imaginary axis we must
specify a class of functions by demanding analyticity (except in the
branching points). In \cite{prd57} it was the analyticity requirement that
enabled us to find the mass spectrum in the limit of large black holes.
The spectrum depends on two quantum numbers. This is explained by the fact
that we succeeded in accounting for the motion of the shell in all
regions of Kruskal space-time. The configuration space variable, the
radius $R$ of the shell takes its values on the cross (see Fig.\ 
\ref{penrose}) with intersection at the horizon point $R=R_g$ ($R_g$ is
the gravitational radius of the shell). $R$ changes from $0$ to $R_g$ on
two of the hands of the cross in $T_\pm$ regions of the space-time and
 from $R_g$ to $\infty$ on the other two hands in $R_\pm$
regions of Kruskal space-time. So the principal difference from the usual
one dimensional quantum mechanics, where the particle moves along a real
line, $R^1$ was that the configuration space was not even two copies of a
real line but some nontrivial topological space (a cross) which is not
even a manifold.  The nontrivial topology of the configuration space
results in an appearance of the second quantum number.

In this paper  we consider the quasiclassical solutions of the
finite-difference Schroedinger equation of the quantum black hole model
considered in \cite{prd57}.  The Riemannian surface
on which the finite difference Schroedinger equation
is defined turns out to be a sphere with two punctured points which are
two spatial infinities  at $R_+$ and $R_-$ regions of Kruskal
space-time (Fig.\ \ref{penrose}).  There is a nontrivial cycle on this
Riemannian surface. The usual Bohr-Sommerfeld condition, that corresponds
to the bound motion of the shell, gives one of the quantum numbers in
the spectrum. Apart from this condition  there appears to exist another
nontrivial quantization condition which stems from the requirement for the
wave function to be a regular function on the relevant Riemannian surface.
It therefore should have trivial monodromy along the nontrivial cycle on
the Riemannian surface.  This requirement leads to another
Bohr-Sommerfeld-like condition on the mass spectrum. So the quasiclassical
mass spectrum depends on two quantum numbers.  It becomes clear in our
considerations that the appearance of the second quantum number is due to
the nontrivial topology of the classical configuration space of the model.

We will remind the black hole model considered in \cite{prd57} in sections
\ref{sec:model} and \ref{sec:dirac} and write the finite difference
Schroedinger equation which we will analyze (it is slightly different from
the one considered in \cite{prd57} because of another factor ordering).
Then the method of construction of quasiclassical solutions of
differential equations in complex domain will be explained in section\
\ref{sec:fedor}. We show how  Bohr-Sommerfeld
quantization condition arises from gluing the global quasiclassical
solution from solutions defined in different regions of the complex plane.
In sections\ \ref{sec:surface} and\ \ref{sec:bh}  we apply this method for
the Schroedinger equation of our black hole model and obtain the
quasiclassical mass spectrum which posses the special properties mentioned
above. It is different from the large black hole spectrum obtained in
\cite{prd57} but coincides with it in a certain limit.  In section\
\ref{sec:collapse} we show that one of the quantization conditions remains
valid for the unbound motion of the shell and therefore defines the
spectrum of the states of the  collapsing shell. In the case of a
null-dust shell when the continuous parameter $M$ -- bare mass of the
shell is absent, this quantization condition leads to discrete mass
spectrum for unbound motion.  This spectrum turns out to be $m\sim m_{pl}
\sqrt{k}$, $k\in Z$ -- the spectrum found by Bekenstein and Mukhanov
\cite{bekenstein}.

\section{Hamiltonian dynamics of the selfgravitating thin shell.}
\label{sec:model}


We consider the following  model of the black hole \cite{prd57}. This is a
self-gravitating spherically symmetric dust thin shell, endowed with a
bare mass $M$. The whole space-time is divided into three different
regions: the inner part ( I ), the outer part ( II ) both containing no
matter fields and separated by a thin layer ( III ), that contains the
dust matter of the shell ( Fig.\ \ref{space} where the Carter-Penrose
diagram of the space-time is presented.).  The general metric of a
spherically symmetric space-time is:
  \begin{equation}
  \label{metric}
ds^2=-
N^2 dt^2 + L^2 ( dr + N^r dt )^2 + R^2 ( d\theta^2 + \sin^2\theta d\phi^2
)
 \end{equation}
where $(t,r,\theta ,\phi )$ are space-time coordinates, $
N, N^r, L, R$ are some functions of $t$ and $r$ only. The trajectory of the
thin shell is some 3-dimensional surface $\Sigma $ in the space-time
given by some function $\hat r(t)$:  $\Sigma^3=\{ (t,r,\theta ,\phi ):
r=\hat r(t)\} $.  In the region I $ r< \hat r-\epsilon $, in the  region
II $ r>\hat r+\epsilon $, and the region III is a thin layer $\hat
r-\epsilon <r <\hat r+\epsilon $.  We require that metric coefficients
$N,N^r,L$ and $R$ are continuous functions, but jump discontinuities could
appear in their derivatives at the points of $\Sigma $ when the limit
$\epsilon\rightarrow 0$ is taken.  The action functional for the system of
the  spherically symmetric gravitational field and the thin shell is
 \begin{eqnarray}
 \label{action}
 S=S_{gr}+S_{shell}=\frac{1}{16\pi
G}\int\limits_{I+II+III}\ ^{(4)}R\sqrt{-g} d^4x \nonumber\\+ (surface\
terms) - M\int\limits_{ \Sigma } d\tau
 \end{eqnarray}
It consists of the standard Einstein-Hilbert action for the gravitational
field and the matter part of the action that describes the  thin shell of
dust.

The complete set of degrees of freedom of our system consists of the set
of $N(r,t), N^r(r,t), L(r,t), R(r,t)$ which describe the gravitational
field and $\hat r(t)$ which describes the motion of the shell.  The metric
(\ref{metric}) has the standard ADM form for 3+1 decomposition of a
space-time with lapse function $N$ , shift vector $N^i=(N^r,0,0)$ and
space metric $h_{ik}=diag( L^2, R^2, R^2\sin^2\theta)$.

Substitution of the expression (\ref{metric}) for the metric into the
action (\ref{action}) gives
in Hamiltonian formalism \cite{prd57}
 \begin{eqnarray}
 \label{hamil}
 S=\int\limits_{I+II}\left(
P_L\dot L+P_R\dot R-NH-N^rH_r\right) dr dt+ \int\limits_\Sigma
\hat{\pi}\dot{\hat{ r}}- \nonumber\\ \hat N \left( \hat R\left[ R'\right]
/(G\hat L) + \sqrt{m^2+ \hat\pi^2/ \hat L^2} \right) - \nonumber\\ \hat
N^r \left(-\hat L\left[ P_L\right]-\hat\pi\right) dt
 \end{eqnarray}
where $P_R(r), P_L(r)$ are momenta conjugate to $R(r)$ and $L(r)$, $\pi$
is momentum conjugate to $\hat r$, hats denote the variables defined on
the shell surface $\Sigma$, dots denote time derivatives $\partial
/\partial t$ and primes denote derivatives $\partial /\partial r$.
Square brackets denote the jump of a function on the shell surface:
$\left[ {\cal A} \right] =\mbox{lim}_{\epsilon\rightarrow 0}\left(
{\cal A}(\hat r+\epsilon )-{\cal A}(\hat r-\epsilon )\right)$. The lapse
$N(r)$ and $\hat N$, the shift $N^r(r)$ and $\hat N^r$ functions become
the Lagrange multipliers as usual in ADM formalism and $H$ and $H^r$ are
the constraints:
 \begin{equation}
 \label{constraints}
 \left\{
 \begin{array}{rcl}
H&=& G\left(
\frac{\displaystyle LP_L^2}{\displaystyle 2R^2}- \frac{\displaystyle
P_LP_R}{\displaystyle R}\right) +\\ & & \frac{\displaystyle
1}{\displaystyle G}\left( - \frac{\displaystyle L}{\displaystyle 2}-
\frac{\displaystyle (R')^2}{\displaystyle 2L}+ \left( \frac{\displaystyle
RR'}{\displaystyle L}\right)'\right) \\ H_r&=&P_RR'-LP_L'.
 \end{array}
 \right.
 \end{equation}
The system of constraints contain two surface
constraints in addition to usual Hamiltonian and momentum constraints of
the ADM formalism.

ADM constraints:
 \begin{equation}
 \label{constraints1}
\left\{
\begin{array}{l}
H=0\\
H_r=0\\
 \end{array}
 \right.
 \end{equation}

Shell constraints:
 \begin{equation}
 \label{shellconstraints}
\left\{
\begin{array}{l}
\hat H_r=\hat\pi+\hat L\left[ P_L\right] =0\\
\hat H=\frac{\displaystyle R\left[ R'\right] }{\displaystyle GL}+
\sqrt{M^2+\left.\hat\pi^2\right/ L^2}=0
 \end{array}
 \right.
 \end{equation}

Karel Kuchar \cite{kuchar} proposed some specific canonical transformation
of the variables $(R, P_R, L, P_L)$ to new canonical set $(R, \bar P_R, m,
P_m)$ in which Hamiltonian and momentum constraints given by
(\ref{constraints}) are equivalent to the very simple set of constraints :
 \begin{equation}
 \label{simple}
 \left\{
 \begin{array}{l}
\bar P_R=0\\  m'=0
 \end{array}
 \right.
 \end{equation}
The idea is to use the Schwarzschild ansatze
for the space-time metric instead of the metric (\ref{metric}):
 \begin{eqnarray}
 \label{schwar}
ds^2=-F(R,m) dT^2+\frac{\displaystyle
1}{\displaystyle F(R,m)} dR^2+\nonumber\\ R^2 (d\theta^2+\sin^2\theta
d\phi^2)
 \end{eqnarray}
where $T, R$ and $m$ are some functions of $(r,t)$ and
$F(R,m)=1-\left. 2Gm \right/ R$.
One could construct a canonical transformation between   $(R,
P_R, L, P_L)$ and
 $(R, \bar P_R, m, P_m)$ so that the system of constraints
(\ref{constraints1}) is equivalent to the system of constraints
(\ref{simple}) in new variables in the phase space.

In the presence of the thin shell the configuration space
also contains the coordinates $\hat R, \hat L$ and $\hat r$.
If we introduce the coordinates
 \begin{equation}
 \label{newtransf}
 \begin{array}{rcl}
\hat p&=&\hat\pi+L\left[ P_L\right]\\
\\
\hat P_{\hat R}&=& \pm \left[ \frac{\displaystyle 1}{\displaystyle 2G}
R\ln{\left| \frac{\displaystyle RR'-GLP_L}{\displaystyle
RR'+GLP_L}\right|} \right]
 \end{array}
 \end{equation}
on the shell surface we could see that these coordinates turn out to be
conjugate to $\hat R$ and $\hat r$ \cite{prd57}.
The set $(m(r), P_m(r),
R(r), \bar P_R(r),\hat R, \hat P_{\hat R}, \hat r, \hat p)$ gives the
canonical coordinates in the whole phase space of the system. One could
consider the whole set of constraints (\ref{constraints1}) and
(\ref{shellconstraints}) in the phase space $\Pi=\{ (R(r,t), P_R(r,t),
L(r,t), P_L(r,t), \hat {r}(t), \hat{\pi}(t)) \}$. The surface momentum
constraint $\hat H_r=0$ (\ref{shellconstraints}) takes the form
\begin{equation} \label{exclude} \hat p=0 \end{equation} The shell
Hamiltonian constraint is expressed through the variables as follows
\widetext
 \begin{equation}
 \label{hamilton}
\hat H=
\frac{\displaystyle R}{\displaystyle G}
\sqrt{\sqrt{F_{out} }-\sqrt{F_{in}}  \exp \left( \frac{\displaystyle G\hat P_{\hat R}}{\displaystyle R}
\right)}
\sqrt{\sqrt{F_{out}}-\sqrt{F_{in}}  \exp \left( -\frac{\displaystyle G\hat P_{\hat R}}{\displaystyle R}
\right)} - M=0
 \end{equation}
\narrowtext
which means that
 \begin{equation}
 \label{exponent}
\exp \left( \pm\frac{\displaystyle G\hat P_{\hat R}}{\displaystyle R}
\right) =
\frac{\displaystyle 1}{\displaystyle 2\sqrt{F_{in}F_{out}}}
\left(
 F_{in}+F_{out}-
\frac
{\displaystyle M^2G^2}
{\displaystyle R^2}
\pm
\frac
{\displaystyle 2MG}
{\displaystyle R}
{\cal Z}
\right)
 \end{equation}
where

 \begin{equation}
 \label{z}
{\cal Z}=\sqrt{\left( F_{out}-F_{in}\right)^2\frac{\displaystyle R^2}{
\displaystyle 4 M^2G^2}-\frac{1}{2}\left( F_{out}+F_{in}\right)
+\frac{\displaystyle M^2G^2}{\displaystyle 4 R^2}}
 \end{equation}

Since the shell could be found in each of the four regions $R_\pm,
T_\pm$ of Kruskal space-time the dynamical variable $\hat R$ which
measures the radius of the shell $\Sigma$ embedded in the space-time $M$
could take its values on the cross, shown on Fig.\ \ref{penrose}. So the
configuration space of the dynamical system under configuration has the
singularity at the horizon. We will see how this singularity disappears
when we turn to the quantum mechanics where we'll have to consider the
realization of the relevant operators on some complex Riemannian surface
(section \ref{sec:bh}).

The Hamiltonian constraint (\ref{hamilton}) was derived under the assumption
that both $F_{in}$ and $F_{out}$
are positive. It is possible to derive analogous
constraints in $T_\pm$-regions, where $F<0$. But, instead, one could
consider the function $F^{1/2}$ as a complex valued function. The
point of the horizon $F=0$ becomes a branching point , and we need the
rules of the bypass. We assume the following
 \begin{equation}
 \label{bypass2}
 \begin{array}{rl}
F^{1/2}&=\left| F\right| e^{i\phi}\\ \phi =0& \mbox{\
in\ } R_+\mbox{-region}\\ \phi = \pi /2& \mbox{\ in\ } T_-\mbox{-region}\\
\phi =\pi & \mbox{\ in\ } R_-\mbox{-region}\\ \phi =-\pi /2& \mbox{\ in\ }
T_+\mbox{-region}\\
 \end{array}
 \end{equation}
The reason for  such
analytical continuation is that we will be able to get the single
equation on the wave function $\Psi$ which covers all four patches of the
complete Penrose diagram for the Schwarzschild space-time. Some important
consequences of this fact will become evident in section \ref{sec:bh}.

The case of special interest for us will be the dynamics of the null-dust
shell which corresponds to the case $M=0$ so that the shell propagates
with the speed of light \cite{louko}. In this case the shell constraints
(\ref{shellconstraints}) take a simple form
 \begin{equation}
 \label{}
\left\{
\begin{array}{l}
\hat H_r=\hat\pi+\hat L\left[ P_L\right] =0\\
\hat H=\frac{\displaystyle R\left[ R'\right] }{\displaystyle GL}+
\hat\pi=0
 \end{array}
 \right.
 \end{equation}
and from (\ref{hamilton}) we get the form of shell constraint in terms of
Kuchar variables:
 \begin{equation}
 \label{iso}
\frac{\displaystyle R^2}{\displaystyle G^2}\sqrt{
F_{in}+F_{out}-2\sqrt{F_{in}F_{out}}\mbox{ ch}\left(\frac{G\hat P_{\hat
R}}{\hat R}\right)}=0
 \end{equation}
which is equivalent to
 \begin{equation}
 \label{iso1}
\exp\left( \pm\frac{\displaystyle G \hat P_{\hat
R}}{\displaystyle R} \right)=\sqrt{\frac{\displaystyle F_{in}}
{\displaystyle F_{out}}}
 \end{equation}

In the rest of the paper we will restrict ourselves with the motions of the
shell when $m_{in}=0$.
In this case it is convenient to make a canonical transformation from
 $(\hat R, \hat P_{\hat R})$
to $(\hat S, \hat P_S)$:
\begin{equation}
\label{area}
\left\{
\begin{array}{rcl}
\hat S&=&\frac{\displaystyle \hat R^2}{\displaystyle
(2Gm)^2}=\frac{\displaystyle \hat R^2}{\displaystyle R_g^2}\\ \hat
P_S&=&R_g^2\frac{\displaystyle \hat P_{\hat R}}{\displaystyle 2R}
\end{array} \right.  \end{equation} where $R_g$ is the gravitational
radius of the shell. Dimensionless variable $\hat S$ is the surface area
of the shell measured in the units of the horizon  area of the shell of
mass $m$.

\section{Dirac Quantization of the model.}
\label{sec:dirac}

The Dirac quantization of the black hole model under consideration
looks like the follows.

The phase space of our model consists of
coordinates $(R(r), \tilde P_R(r), m(r),\\
 P_m(r), \hat R, \hat P_{\hat R}, \hat r, \hat p_r)$
$r\in (-\infty, \hat r-\epsilon )\bigcup (\hat r+\epsilon, \infty)$.
Then the wave function in coordinate representation depends on configuration space coordinates:
\begin{equation}
\label{wave}
\Psi=\Psi(R(r), m(r), \hat R, \hat r)
\end{equation}
and all the momenta become operators of the form
 \begin{eqnarray}
\label{operators}
\tilde P_R(r)=-i\left.\delta\right/\delta R(r); &\
P_m(r)=-i\left.\delta\right/ \delta m(r);\nonumber\\ \hat P_{\hat
R}=-i\left.\partial\right/\partial \hat R; & \hat
p_r=-i\left.\partial\right/\partial \hat r.
 \end{eqnarray}
Using the Kuchar constraints (\ref{simple}) and the shell
constraint (\ref{exclude}) in operator form we conclude that the wave
function does not depend on $R(r)$ and $\hat r$ as far as
 \begin{equation}
 \label{qkuchar}
 \left\{
 \begin{array}{rcl}
\left.\partial \Psi\right/\partial R(r)&=&0\\
m'(r)\Psi&=&0\\
\left.\partial \Psi\right/\partial \hat r&=&0
 \end{array}
 \right.
 \end{equation}
The dependence on $m(r)$ is
reduced in regions I and II to $\Psi\equiv \delta(m-m_{\pm})$ where
$m_{\pm}$ do no depend on $r$. $m_\pm$  equal
to Schwarzschild masses in the inner and outer regions $m_{in}$ and
$m_{out}$. We restrict ourselves with the case when $m_{in}=0$ so the
dependence of the wave function on the phase space variables reduces to
 \begin{equation}
\Psi =\Psi (m, \hat R)
 \end{equation}
The only nontrivial equation is the shell hamiltonian
constraint (\ref{hamilton}). It contains the square root expression which
is difficult to realize at quantum level. So we use the squared version of
the shell Hamiltonian constraint, which in terms of the canonical pair
$(S, P_s)$ (\ref{area}) reads as
  \begin{equation}
\label{Cons} \hat
C=1-\frac{\displaystyle 1}{\displaystyle 2\sqrt{S}}-\frac{\displaystyle
M^2}{\displaystyle 8 m^2S}-\sqrt{F}\mbox{ ch}\left(\frac { \hat
P_S}{2Gm^2}\right) =0
 \end{equation}

The operator $\hat C$ contains the
exponent of the of the momentum $\hat P_S$.  This exponent becomes an
operator  of finite displacement when $\hat P_S$ is the differential
operator:
 \begin{equation}
 \label{finite}
\exp\left( \frac{\hat P_S}{2Gm^2}\right) \Psi=
\exp\left( -i\frac{m^2_{pl}}{2 m^2}\frac{\partial}{\partial \hat
S}\right) \Psi = \Psi(\hat S -\zeta i)
 \end{equation}
where $m_{pl}$ is Plank mass and $\zeta =\frac {1}{2} (\frac {m_{pl}}{
m})^2 $.

The constraint $\hat C$ becomes an equation in finite differences  if we
substitute the expression (\ref{finite})
into  (\ref{Cons}).
 \begin{equation}
 \label{main}
 \Psi (m, S+i\zeta )+\Psi (m, S-i\zeta )=F^{-1/2}
\left( 2-\frac{\displaystyle 1}{\displaystyle
\sqrt{S}}-\frac{\displaystyle M^2}{ \displaystyle 4m^2S}\right)
\Psi (m,S)
 \end{equation}
The shift in the argument of the wave function is along an imaginary axis.
In the case of differential equation we require the solution to be
differentiable sufficiently many times. Similarly, we have to demand the
solutions of our finite differences equation (\ref{main}) to be analytical
functions.  This condition is very restrictive but unavoidable. The
importance of this requirement is shown in \cite{shell} where it is the
analyticity of the wave functions and not the boundary conditions that
lead to the existence of the discrete mass (energy) spectrum for bound
states. In the next section we will see how it works in the quasiclassical
regime.

The construction of quasiclassical solutions of the differential (finite
difference) equations in complex domain requires the use of a special
technic \cite{fedoruk} explained in the next section. It is a well known
fact that the quasiclassical approximation is not valid in the vicinity of
the turning points of the classical trajectories of the system. When
solving the equation in complex domain one could use the quasiclassical
ansatze for the approximate solution only in regions that are distant from
the turning points on the complex plane \cite{ll}. One need to glue the
global solution defined on the whole complex plane from the solutions
defined in different regions. This  global solution must be an
approximation to some analytical solution of the differential equation
under consideration.

The next section is devoted to the explanation of the method of
constructing the quasiclassical solutions of a differential equation in
complex domain \cite{fedoruk} using a simple example of nonrelativistic
Schroedinger equation for the particle moving in a potential well. Then in
section \ref{sec:bh}  we use this method in order to build the analytical
quasiclassical solutions of (\ref{main}) in complex domain which will
enable us to find the quasiclassical mass spectrum of our black hole
model.

\section{Quasiclassical solutions of Schroedinger equation in complex
domain.}
\label{sec:fedor}

In this section we explain the construction of quasiclassical
approximation to a regular solution of Schroedinger equation on the
complex plane. The method was developed in \cite{fedoruk} and
\cite{heading} where all necessary theorems could be found.

Let us consider the nonrelativistic Schroedinger equation
 \begin{equation}
 \label{SCHR}
-\hbar^2\Psi''(z)+q(z)\Psi (z)=0
 \end{equation}
on the complex plane $z\in C$. Let the function $q(z)$ be holomorphic in
a region $D\in C$.  Then in accordance with Cauchy theorem
\cite{tfkp} equation (\ref{SCHR}) has a solution regular in $D$.

We look for the approximate solutions of (\ref{SCHR}) in the form of
quasiclassical ansatze
 \begin{equation}
 \label{ansatze}
\Psi (z) =\exp\left(
\frac{\displaystyle i}{\displaystyle \hbar} \Omega (z)
\right) \sum_{k=0}^{\infty}\hbar^k\phi_k
 \end{equation}
Then substituting (\ref{ansatze}) in (\ref{SCHR}) we obtain in zero order
on $\hbar$ the Hamilton-Jacobi equation on $S(z)$:
 \begin{equation}
 \label{HJ}
\left( \frac{\displaystyle \partial \Omega }{\displaystyle \partial
z}\right)^2 + q(z)=0
 \end{equation}
If $q(z)\not= 0$ in the neighborhood $U_{z_0}$ of some point $z_0$ then
(\ref{HJ}) has two solutions corresponding to the two different branches
of the function $q^{1/2}(z)$:
 \begin{equation}
\Omega (z_0,z)=\pm\int\limits_{z_0}^z\sqrt{-q(t)} dt
 \end{equation}
and
 \begin{eqnarray}
 \label{ccc}
\Psi (z)=a\Phi_1(z)+b \Phi_2(z)\nonumber\\
\Phi_{1,2} (z)=\left( q(z) \right)^{-1/4}
\exp\left( \pm\frac{\displaystyle i}{
\displaystyle\hbar }\int\limits_{z_0}^z\sqrt{-q(t)} dt\right)
 \end{eqnarray}


The coefficients of (\ref{SCHR}) are analytical functions in the
neighborhood of $z_0$. Hence in accordance with Cauchy theorem
in the neighborhood $U_0$ of a point $q(z_0)=0$ exist two regular
solutions of (\ref{SCHR})  as well. Nevertheless,
if we consider the equation (\ref{SCHR}) far enough
from $z_0$ where the quasiclassical approximation is valid, we find
that the  quasiclassical ansatze is not a good
approximation for the regular solution in the punctured neighborhood of the
point $z_0$ (the branching point for the function $q^{1/2}(z)$).  This
could be easily seen if one note that $\Phi_1$ transforms into $i\Phi_2$
after the analytical prolongation along a closed path $\gamma_1$ which goes
around $z_0$ (Fig.  \ref{schro}) while the regular solution have the
trivial monodromy along this path.  The approximate expression (\ref{ccc})
for the wave function is valid only in a certain sector $\alpha
<\mbox{Arg} (z-z_0)<\beta$ in the neighborhood of $z_0$. The
explanation of the effect is the following.

Let $z=z_0$ be, for example, a simple zero of the function $q$ : $q\approx
q_0(z-z_0)$ in $U_0$ ($q_0=\rho_0 e^{i\phi_0}$ and $(z-z_0)=\rho
e^{i\phi}$).  Then
 \begin{equation}
\Omega
(z,z_0)=\frac{\displaystyle
2}{\displaystyle 3} q_0^{1/2}(z-z_0)^{3/2}=\frac{\displaystyle
2}{\displaystyle 3}\sqrt{\rho_0}\rho^{3/2}
e^{i(\phi_0+3\phi -\pi ) /2}
 \end{equation}
This function grows when $\rho =|(z-z_0)|\rightarrow\infty$ in the sector
$-\pi < (3\phi +\phi_0-\pi )<\pi$ confined by the lines $l_1$ and
$l_2$ ( Fig. \ref{schro} where $\phi_0$ is taken to be $\phi_0=\pi$),
then it is pure oscillating on $l_1$, $l_2$ and $l_5$ and is decreasing in
sectors $\pi < (3\phi +\phi_0-\pi )<3\pi$ and $-\pi < (3\phi
+\phi_0)<-3\pi$.  So the solution $\Phi_1$  (\ref{ccc}) grows in
the sector {\bf 1} and decreases in the sectors {\bf 2} and {\bf 3} on
Fig.  \ref{schro}.  The solution $\Phi_2$ has an opposite behavior:  it
grows in sectors {\bf2} and {\bf 3} and decreases in sector {\bf 1}.
Therefore in sector {\bf 1} the solution $\Phi_1$ is exponentially large
compared to $\Phi_2$. But in quasiclassical expression for the wave function we can not retain the
exponentially small items simultaneously with exponentially large in the
approximate expression for the wave function and in fact we could set the
coefficient $b$ to be equal to any number if only $a\not= 0$.  The
explanation for the fact that the quasiclassical ansatze fails to be the
correct approximation for the regular solution is that the coefficients
$a,b$ (\ref{ccc}) of expansion of the wave function on the system of
solutions $\Phi_{1,2}$ are different in the sectors {\bf 1,2} and {\bf 3}.
This fact was observed for the first time by Stokes and the phenomenon is
called the Stokes phenomenon \cite{heading}. The lines $l_i$ defined by
the equation
 \begin{equation}
\mbox{ Im } \Omega (z,z_0)=0
 \end{equation}
are called the Stokes lines. They start at the
branching points $q(z_0)=0$ and terminate at singular points of
equation or at infinity. Both solutions $\Phi_{1,2}$ (\ref{ccc}) are
oscillating along these lines.


When the point $z_0$ is a zero of $n$-th order for the function $q(z)$
( $q(z)\approx q_0(z-z_0)^n$ ) the Stokes lines approach the rays $\mbox{
Arg }(z-z_0)=\phi_k=const$ near $z_0$.  $\phi_k$ are given by the
formula
 \begin{equation}
 \label{angles}
\phi_k=\frac{\displaystyle (2k+1)\pi -\phi_0}{\displaystyle n+2}
 \end{equation}
where $\phi_0=\mbox{ Arg }(q_0)$. The formula remains valid when $n<0$ and
$z_0$ is a singular point of (\ref{SCHR}).

The global structure of the Stokes lines when $q(z)$ has the form of potential
well with two turning points $q(z_{0,1})=0$ is shown on Fig. \ref{schro}. The
Stokes lines divide the complex plane into the  sectors {\bf 1,2,3} and
{\bf 4} (Fig.  \ref{schro}) in which one of solutions (\ref{ccc}) is
exponentially small compared to the other.

Let the approximate expression for the regular  solution be given in one
of the sectors, for example in {\bf 1}
 \begin{equation}
 \label{bbb} \Psi
(z) =a_1\Phi_1(z)+b_1\Phi_2(z),\ z\in\mbox{ {\bf 1} }
 \end{equation}
The form of approximate solution in other sectors
of the complex plane could be found using the using  the
following algorithm \cite{fedoruk}.

1) The whole complex plane
is divided into the set of intersecting regions $\{ D_i\}$ called the
canonical regions.  The boundary of each  region $D_i$ consists of
some Stokes lines $\partial D_i=\bigcup l_k$ so that each $D_i$ is a sum
of the sectors {\bf 1,2,...} into which the Stokes lines divide the complex
plane.  
For example, for the problem of a particle in a  potential well (Fig. 
\ref{schro}) some of canonical regions are $D_1=\mbox{ {\bf 1} }\bigcup
\mbox{ {\bf 2} }$, $D_2=\mbox{ {\bf 3} }\bigcup
\mbox{ {\bf 3} }$ and  $D_3=\mbox{ {\bf 2} }\bigcup
\mbox{ {\bf 4} }$.
The regions $D_i$ are foliated by
the level lines $L_c=\{\mbox{ Im}(\Omega (z))=c=const\}$. Each $D_i$ is
chosen so that $c=\mbox{ Im}(\Omega )$ changes from $-\infty$
to $+\infty$ in $D_i$.  In each of these regions the fundamental system of
solutions $\Phi^{l_i}_{1,2}(z)$
 \begin{eqnarray}
 \label{ddd}
 \Psi (z) =a_1\Phi^{l_i}_1(z)+a_2\Phi^{l_i}_2(z),\ z\in D_i\nonumber\\
\Phi^{l_i}_{1,2}=\frac{\displaystyle 1}{\displaystyle
q^{1/4}(z)}\exp\left(\pm \frac{\displaystyle i}{\displaystyle
\hbar}\Omega^{l_i} (z)\right)
 \end{eqnarray}
is introduced. In order to define the fundamental system of solutions one
need to choose a branching point $z_0\in\partial D_i$ and a Stokes line
$l_i\in D_i$, $z_0\in l_i$. Then the phase $\Omega$ of the quasiclassical
wave function (\ref{ansatze}) is
 \begin{equation} \Omega^{l_i}
(z,z_0)=\int\limits^z_{z_0}\sqrt{-q(t)} dt +\Omega_0
 \end{equation}
where
$\Omega_0$ is chosen so that $\mbox{ Im }(\Omega^{l_i})=0$ on $l_i$.
Assume the first solution of the fundamental system is increasing in the
sector on the right (clockwise) from the reference Stokes line and is
decreasing in the sector on the left (anti-clockwise) from the reference
Stokes line. 

2) In intersections $D_i\bigcap D_j$ the transition matrices $T_{i,j}$
between the corresponding systems of solutions $\Phi^{l_i}_{1,2}$ and
$\Phi^{l_j}_{1,2}$ are defined:
 \begin{eqnarray}
 \Psi =a_i\Phi^{l_i}_1+b_i\Phi^{l_i}_2=a_j\Phi^{l_j}_1+b_j\Phi^{l_j}_2,\
z\in D_i\bigcap D_j \nonumber\\ \left( \begin{array}{c} a_i \\ b_i
\end{array} \right) = T_{i,j}  \left( \begin{array}{c} a_j \\ b_j
\end{array} \right)
 \end{eqnarray}

3) Let the form of the solution be given in some region $D_{1}$. Then we
could find the form of the quasiclassical solution at any point $A$ of the
complex plane. We need to draw a path $\gamma$ from a point $B\in D_{1}$
to $A$ and to cover it by the set of intersecting canonical regions $D_1,
...,D_k$.  The form of the solution in the region $D_k$,  which contain
the point $A$, is
 \begin{eqnarray}  
 \Psi (z)=a_k\Phi^{l_k}_1+b_k\Phi^{l_k}_2,\ z\in D_k \nonumber\\
\left( \begin{array}{c} a_k \\ b_k \end{array} \right) =T
\left( \begin{array}{c} a_1 \\ b_1 \end{array} \right)
 \end{eqnarray}
where the matrix $T$:
 \begin{equation}
T=T_{k,k-1}\times T_{k-1,k-2}\times ...T_{2,1}
 \end{equation}

Let us consider the quasiclassical solutions of equation (\ref{SCHR}) in
the simplest case when the function $q(z)$ has two zeros (the turning
points of the classical motion) situated in points $z_0$ and $z_1$ on real
line (see Fig. \ref{schro}).  We impose on the wave function $\Psi$ the
requirement $\Psi (z)\rightarrow 0$ when $z\rightarrow \pm\infty$ along
the real line. So the solution in the region $D_1=\mbox{ {\bf 1} }\bigcup 
mbox{ {\bf 2} }$ must be
 \begin{eqnarray}
\Psi (z)=a_1\Phi^{l_1}_1+b_1\Phi^{l_1}_2,\ z\in D_1\nonumber\\
\left( \begin{array}{c} a_1 \\ b_1 \end{array} \right) =
\left( \begin{array}{c} 0 \\ 1 \end{array} \right)
 \end{eqnarray}
since the second solution of the fundamental system in region $D_1$ is
decreasing in the sector {\bf 1} in accordance with the convention taken
for enumeration of the solutions from the fundamental system.  We need to
find the solution in the region $D_3=mbox{ {\bf 2} }\bigcup \mbox{ {\bf 4} }$ 
which contain the real line on the
left from the region of the classical motion
 \begin{equation}
\Psi (z)=a_3\Phi^{l_3}_1+b_3\Phi^{l_3}_2,\ z\in D_3
 \end{equation}
 We choose a path
$\gamma_2$ connecting  points $A\in D_3$ and $B\in D_1$ (Fig. \ref{schro}).
It could be covered by the three canonical regions $D_1$,
$D_2$ and $D_3$  so that
 \begin{equation}
\label{t1}
\left( \begin{array}{c} a_3 \\ b_3 \end{array} \right) =
T\left( \begin{array}{c} 0 \\ 1 \end{array} \right)
 \end{equation}
where
 \begin{equation}
 \label{trans}
T=T_{3,2}\times T_{2,2}\times T_{2,1}
 \end{equation}
The matrix $T_{2,2}$ has appeared in the equation because in the canonical
region $D_2$ there exist two different fundamental systems of solutions.
One of them is defined by the Stokes line $l_2$ and the branching point
$z_0$ while the other is defined by the Stokes line $l_2$ and the
branching point $z_1$ (Fig. \ref{schro}). The transition matrix between
the two fundamental systems of solutions  in $D_2$ is
 \widetext
 \begin{equation}
T_{2,2}=e^{i\epsilon}\left( \begin{array}{cc}
0&\exp\left( \frac{i}{\hbar}\int\limits^{z_1}_{z_0}\sqrt{-q(z)} dz
\right) \\ \exp\left(
-\frac{i}{\hbar}\int\limits^{z_1}_{z_0}\sqrt{-q(z)} dz \right) &
0\end{array} \right)
 \end{equation}
where $e^{i\epsilon}$ is some factor coming from the normalization of the
solutions, and which is not important for us. The transition matrix between
the fundamental system of solutions defined in the region $D_1$ with
the Stokes line $l_1$ with the turning point $z_0$ and the fundamental
system of solutions defined in the region $D_2$ with the Stokes line $l_2$
and the same turning point $z_0$ is
 \begin{equation}
T_{2,1}=e^{i\pi/6}\left( \begin{array}{cc}
0&1\\ 1 &
i  \end{array}
\right)
 \end{equation}
The same transition matrix is between the region $D_2$ with the Stokes
line $l_2$ and the turning point $z_1$ and the region $D_3$ with the
Stokes line $l_3$ and the turning point $z_1$. In accordance with
(\ref{trans}) we find
 \begin{equation}
 \label{t2}
T= e^{i(\epsilon +\pi/3)}\left( \begin{array}{cc}
\exp\left( \frac{i}{\hbar}\Omega
(z_0,z_1)\right) &i \exp\left( \frac{i}{\hbar}\Omega
(z_0,z_1)\right) \\ 0 & i\left(
\exp\left( \frac{i}{\hbar}\Omega
(z_0,z_1)\right) +
\exp\left( -\frac{i}{\hbar}\Omega
(z_0,z_1)\right)\right) \end{array} \right)
  \end{equation}
\narrowtext
where $\Omega (z_0,z_1)=\int\limits^{z_1}_{z_0}\sqrt{-q(z)} dz$.
Since we impose the requirement that the quasiclassical wave function
decreases in the sector {\bf 4}, then the expansion of the wave function on
the fundamental system of solutions in region $D_3$ must be
 \begin{equation}
\left( \begin{array}{c} a_3 \\ b_3 \end{array} \right) =
\left( \begin{array}{c} 1 \\ 0 \end{array} \right)
 \end{equation}
because the first solution of the fundamental system defined by the Stokes
line $l_3$ decreases on the left from $l_3$. From (\ref{t1}) and (\ref{t2})
we see that this requirement is satisfied only if
 \begin{equation}
\exp\left( \frac{i}{\hbar}\int\limits^{z_1}_{z_0}\sqrt{-q(z)} dz
\right)=\pm i
 \end{equation}
This condition is equivalent to the well known Bohr-Sommerfeld
quantization condition
 \begin{equation}
 \label{bz}
\frac{\displaystyle 1}{\displaystyle
\pi\hbar}\int\limits^{z_1}_{z_0}\sqrt{-q(z)} dz=n+\frac{\displaystyle
1}{\displaystyle 2},\ n\in Z
 \end{equation}

So we have found that in the simplest case when the regular function
$q(z)$ from the Schroedinger equation (\ref{SCHR}) has just two
simple zeros $q(z)=0$ situated on the real line, the quasiclassical
solution defined on the whole complex plane exists if the coefficients of
the equation satisfy Bohr-Sommerfeld condition (\ref{bz}). The approximate
expression for the solution of the form (\ref{ansatze}) is different in
the different canonical regions $D_i$ of the complex plane and the
approximation itself is valid only far from the turning points $q(z)=0$ of
the equation (\ref{SCHR}).

\section{Quasiclassical Solutions of Schroedinger equation for
selfgravitating dust shell}
\label{sec:surface}


In this section we will find the quasiclassical solutions in complex domain
for the equation (\ref{main})
which is the quantum form of Hamiltonian constraint
for the system that consists of a selfgravitating thin dust shell and
its own gravitational field.

A principal difference of the equation (\ref{main}) from the
nonrelativistic Schroedinger equation (\ref{SCHR}) that was considered in
the previous section is that its coefficients contain the function
 \begin{equation}
 \label{F}
F^{1/2}=\sqrt{1-\frac{1}{\rho}}=\frac{\sqrt{\rho (\rho -1)}}{\rho}
 \end{equation}
(where $\rho =\sqrt{S}$) that is a branching analytical
function on the complex plane. So we have to consider the equation
(\ref{main}) not on the complex plane but instead on the Riemannian
surface $S_F$ on which the coefficients of the equation are regular
functions.


The Riemannian surface for the function $F^{1/2}$ (\ref{F}) is a
two-dimensional sphere consisting of two Riemannian spheres $S_+$ and
$S_-$  glued along the sides of the cuts  made
in both spheres along the interval $(0,1)$.
 Let us consider the real section $\mbox{Im}(\rho
)=0$ of the Riemannian surface $S_F$. Its part $\rho >0$ turns out to be
a cross with two ends situated in the points $\rho =+\infty$ of $S_F$
(see Fig. \ref{penrose}).
so
that $F^{1/2}>0$ in $\rho =+\infty$ in $S_+$ ( $\infty$ at $R_+$ domain )
and
 $F^{1/2}<0$ in $\rho =+\infty$ in $S_-$ ( $\infty$ at $R_-$ domain ). Two
other ends are in the points $\rho =0$ and the argument of $F^{1/2}$ is
$+\pi /2$ ( $T_-$ domain) in one of the ends and $-\pi /2$ ( $T_+$
domain ) in the other.  So we could naturally identify the cross in the
real section of the Riemannian surface $S_F$ with the cross discussed in
the  section \ref{sec:model} which is the classical configuration space of
the dynamical system under consideration.

An important feature of the form of Hamiltonian constraint for the black
hole model introduced in \cite{prd57} is that by ascribing the special
choice of the argument for the function $F^{1/2}$ as a complex valued
function one could obtain the constraint which is valid in all the four
regions of Kruskal space-time  $R_+, R_-, T_+$ and $T_-$. The price
for this was that the classical configuration space of the dynamical
system had a singular point at the black hole horizon and the
configuration space turned out to be the cross (which is not even a
manifold). Nevertheless, when we pass to the quantum mechanical problem,
we have to consider the complex Riemannian surface $S_F$ as the
configuration space for the system which is a manifold and, as it will be
shown, the dynamical system has no singular point at the horizon if we
choose the coordinate which is regular on the surface $S_F$ in the
neighborhood of the horizon.

We will construct the quasiclassical wave function of the equation
(\ref{main}) on the whole surface $S_F$.  According to the method
explained in the previous section we need first to find the turning and
singular points of the equation. Then we have to determine the structure
of Stokes lines and canonical regions on the Riemannian surface $S_F$. The
last step is to find the corresponding fundamental system of solutions
in each region and the transition matrices between them.

In the limit of large black holes the
displacement parameter $\zeta=\left. m^2_{pl}\right/ 2m^2$ becomes small
and we could cut the Tailor expansion of $\Psi (S+i\zeta )$ on the second
term \cite{prd57}. Then equation (\ref{main}) becomes an ordinary
differential equation:
 \begin{equation}
 \label{ode}
-\zeta^2\Psi
''(S)+\left( 2-F^{-1/2}\left( 2-\frac{\displaystyle 1}{ \displaystyle
\sqrt{S}}-\frac{\displaystyle M^2}{\displaystyle 4m^2S}\right) \right) 
\Psi (S)=0
 \end{equation}
This approximate equation is valid rather far from the horizon ($|\rho |\gg
1$) in both $S_+$ and $S_-$ components of $S_F$. So we expect that the
two equations (\ref{main}) and (\ref{ode}) have common
quasiclassical solutions in the vicinities of $\rho =\infty$ in $S_+$
and $S_-$. We will use this assumption in order to glue the
quasiclassical solutions of (\ref{main}) in different canonical regions
in the neighborhood of infinities.

One important note should be made at this
point.  We have seen in the previous section that the quasiclassical
ansatze could not be valid in the neighborhood of any turning point
($q(z)=0$) of nonrelativistic Schroedinger equation.  Nevertheless, the
Cauchy theorem guaranteed that there exists a regular solution of the
equation in a domain containing this point and we could build its
quasiclassical approximation everywhere except for the neighborhoods of the
turning points. Now the picture is different. We have no theorem
for the equation in finite differences which affirms the existence of
regular solutions of the equation with regular coefficients. So the
assumption we made is in fact the assumption that there exists a solution
of the equation (\ref{main}) regular everywhere on $S_F$ except for the
singular points of the finite difference equation (\ref{main}) (which are
$\rho =\infty$ at $S_+$ and $S_-$ components of $S_F$ as we will see) and
we look for its approximate expression in the form of quasiclassical
ansatze in the regions located far from turning and singular points of
(\ref{main}) on $S_F$.


For the massive dust shell there exist three qualitatively different
types of classical motion, described in \cite{prd57}.\\
a) when $m<M<2m$ the trajectory has the form shown on Fig. \ref{space}a.
It starts from $\rho=0$ in $T_+$ region, crosses its own horizon
$\rho=1$, expands to some maximal radius $\rho_{max}>1$ in $R_+$ region
(it could be observed by an observer at infinity during this period) and
then it collapses to the singularity in the $T_-$ region (the situation
is called "the black hole case" in terms of \cite{prd57}).\\
b) when $M>2m$ the shell is in $R_-$ region at the moment of its maximal
expansion as it is shown on Fig. \ref{space}b, so for the external observer at
$R_+$ infinity it does not appear
from its horizon during the whole evolution  ("the wormhole case").\\
c) the case $M<m$ describes the situation of collapse (Fig.
\ref{space}c) when the trajectory goes from $\rho =\infty$ in $R_+$
region to $\rho =0$ in $T_-$ region.

We will consider in the next section the wave function of the black hole
case (a).

\section{The Quasiclassical Wave Function in the Case of Black Hole.}
\label{sec:bh}

Let us choose the wave function in the form
\begin{equation}
\label{}
\Psi (S)=\exp \left( \frac{i}{\hbar}\Omega (S)\right) (\phi (S)+\hbar \phi
(S)+...)
\end{equation}
of quasiclassical ansatze. Then the phase
$\Omega (S)$ satisfies the Hamilton-Jacobi equation
 \begin{equation}
\label{HJ'}
\mbox{ch } \left( \frac{\partial\Omega}{\partial
S}\right)^2-F^{-1/2}\left( \displaystyle 1-\frac{\displaystyle 1}{ \displaystyle
2\sqrt{S}}-\frac{\displaystyle M^2}{\displaystyle 8m^2S}\right) =0.
 \end{equation}
Its solutions are given by the functions
$\Omega=\int\limits_{S_0}^{S}P_s(s)ds$:
 \begin{equation}
 \label{solution}
P_s=\ln \left(F^{-1/2}\left( 1+\frac{\displaystyle 1}{ \displaystyle
2\sqrt{S}}+\frac{\displaystyle M^2}{\displaystyle 8m^2S}\pm
\frac{\displaystyle M}{\displaystyle 2m\sqrt{S}}{\cal Z} \right) 
\right)
 \end{equation}
where ${\cal Z}$ is given by (\ref{z}).
In terms of the coordinate $\rho$ on the Riemannian surface $S_F$
 \begin{equation}
 \label{rho}
 \left\{
 \begin{array}{rcl}
 \rho &=& \sqrt{S} \\
 P_{\rho} &=& 2\rho P_s
 \end{array}
 \right.
 \end{equation}
we have
 \begin{equation}
P_\rho =2\rho \ln \left(F^{-1/2}\left( 1+\frac{\displaystyle 1}{
\displaystyle 2\rho }+\frac{\displaystyle M^2}{\displaystyle
8m^2\rho^2}\pm \frac{\displaystyle M}{\displaystyle 2m\rho}{\cal
Z}\right) \right).
 \end{equation}

The
turning points $P_s=\left(\left.\partial \Omega \right/\partial
S\right) =0$ are the solutions of the equation
 $$
{\cal Z}=\sqrt{
\frac{\displaystyle m^2}{\displaystyle
M^2}-1+\frac{\displaystyle 1}{\displaystyle
2\rho }+\frac{\displaystyle M^2}{\displaystyle
16 m^2 \rho^2}}=0
 $$


 \begin{eqnarray}
 \label{turning}
 \rho_1=\frac{\displaystyle M^2}{\displaystyle 4m^2(\left.
M\right/ m-1)}, &\mbox{in } R_+ \mbox{region}\nonumber\\
\rho_2=-\frac{\displaystyle M^2}{\displaystyle 4m^2(\left.
M\right/ m+1)}, &\mbox{in } V_- \mbox{region}
 \end{eqnarray}
one of them is in $S_+$ component of $S_F$ (this is the maximum radius of
expansion of the shell)  and the other is in $S_-$ component and is
situated in the region $\rho <0$ which is denoted as $V_-$ on Fig.
\ref{penrose}. In the neighborhood of these turning points $P_s$ is close
to zero and in the equation
 \begin{equation}
\mbox{sh } (P_s) =\frac{\displaystyle {\cal Z} M}{\displaystyle 4m \sqrt{S}
F^{1/2}}
 \end{equation}
we could set $\mbox{sh } (P_s)\approx P_s$ and find that in terms of
regular coordinate $\rho$
 \begin{equation}
P_\rho \sim \sqrt{\rho -\rho_i} ,\ i=1,2
 \end{equation}
so the turning points are just like simple turning points of
nonrelativistic Schroedinger equation considered in the previous section.
In accordance with formula (\ref{angles}) we find that three Stokes lines
originate from each of these turning points as it is shown on Fig.
\ref{stoc2}.


In the neighborhood of infinity in $R_+$ region the regular coordinate is
 \begin{equation}
 \left\{
 \begin{array}{rcl}
t &=& \frac{\displaystyle 1}{\displaystyle \rho}\\
P_t &=& -\frac{\displaystyle P_s}{\displaystyle 2t^3}
 \end{array}
 \right.
 \end{equation}
The solutions of Hamilton-Jacobi equation in the neighborhood of infinity are
 \begin{equation}
P_t=-\frac{\displaystyle 1}{2t^3}\ln\left\{
\frac{\displaystyle 1}{\displaystyle \sqrt{1-t}}\left(
 1-\frac{1}{2}t-\frac{M^2}{8m^2}
t^2\pm\frac{M}{2m}t{\cal Z}\right)\right\}
 \end{equation}
The argument of logarithm is close to one so taking $\mbox{sh }
(P_s)\approx P_s$ we obtain
 \begin{equation}
P_t\approx \pm \frac{i}{8t^2}\sqrt{\frac{M^2}{m^2}-1}
 \end{equation}
So the infinity in $R_+$ region is a singular point of the 4-th order
and from (\ref{angles}) we find that the two Stokes lines
originate from this point as shown on Fig. \ref{stoc3}. This means that
the Stokes lines in the $S_+$ component of the Riemannian surface $S_F$
have vertical asymptotes.

In the neighborhood of infinity point in $R_-$ region Hamilton-Jacobi
equation takes the form
 \begin{equation}
\mbox{ch } \left( -2t^3P_t\right) =-\frac{\displaystyle 1}{\displaystyle 
\sqrt{1-t}}\left(
1-\frac{1}{2}t-\frac{M^2}{8m^2}t^2\right) \approx -1
 \end{equation}
and
 \begin{equation}
 2P_tt^3\approx\pm i\pi;\ 
P_t\sim\frac{\displaystyle 1}{\displaystyle t^3}
 \end{equation}
This singular point is of the 6-th order with four Stokes lines originating
from it, as it is shown on Fig. \ref{stoc3}. The Stokes lines in $S_-$
component of $S_F$ approach asymptotically the lines tilted with the angles
$\phi =\pm\pi/4$ with respect to the real line.

The regular coordinate in the point $\rho =1$ is
 \begin{equation}
 \left\{
 \begin{array}{rcl}
u &=& \sqrt{\rho -1}\\
P_u &=& 4u(1+u^2) P_s
 \end{array}
 \right.
 \end{equation}
In terms of this coordinate
 \begin{equation}
F^{1/2}=\frac{u}{\sqrt{1+u^2}}
 \end{equation}
and
\widetext
 \begin{equation}
 \label{tp}
P_u=4u(1+u^2)\ln\left\{ \frac{\displaystyle 1}{\displaystyle u}\left(
\sqrt{1+u^2}-\frac{1}{2\sqrt{1+u^2}}-\frac{M^2}{8m^2(1+u^2)^{3/2}}
\pm\frac{M}{2m\sqrt{1+u^2}}{\cal
Z}\right) \right\}
 \end{equation}
\narrowtext
The coefficients of the equation (\ref{main}) have singularities in the
point $\rho =1$. But we see from (\ref{tp}) that
$P_u\rightarrow 0$ as $u\rightarrow 0$ and the phase
 \begin{eqnarray}
 \label{faza}
\Omega =\int P_sdS=\int P_udu=-2u^2(\ln u -1/2)+\nonumber\\ (regular\ part)
 \end{eqnarray}
also remains finite as $u\rightarrow 0$. So in terms of the regular
coordinate on the Riemannian surface $S_F$ in the neighborhood of $\rho
=1$ this point is in fact  a turning point ($P_u=0$) of the
quasiclassical solution.


If we consider the truncated equation (\ref{ode}) in terms of regular
coordinate on $S_F$ near the horizon we will see that the point $u=0$ is
the turning point for this equation as well. Indeed, Hamilton-Jacobi
equation for (\ref{ode}) has the form
 \begin{eqnarray}
\frac{P_u^2}{16u^2(1+u^2)^2}=\frac{1}{u}\left( 2\sqrt{1+u^2}-\frac{1}{
\sqrt{1+u^2}}- \right. \nonumber\\
\left. \frac{M^2}{4m^2(1+u^2)^{3/2}}\right)-2
 \end{eqnarray}
and
 \begin{equation}
P_u\sim\sqrt{u}
 \end{equation}
in the neighborhood of $u=0$. So the turning point is just simple turning
point and three Stokes lines originate from it as shown on Fig.
\ref{stoc1}. Note that the $R_+$, $R_-$, $T_+$ and $T_-$ lines intersect
at the horizon as it is shown on Fig.  \ref{stoc1}.
So one of the Stokes lines coincides with $R_+$ while the two others lie
between  $R_-$ and  $T_{\pm}$ lines which means that they
both belong to $S_-$ component of $S_F$.

We are looking for the quasiclassical solution of the finite difference
equation (\ref{main}) in regions far from the turning points. In these
regions the approximate equation (\ref{ode}) is valid. So we expect that
the two equations have the same set of approximate solutions. So we will
use the transition functions between the fundamental systems of solutions
in different canonical regions (see previous section) calculated for the
quasiclassical solutions of (\ref{ode}) in order to glue the
quasiclassical solutions of (\ref{main}) in different canonical regions.
As it was noted at the beginning of the section there is no general
theorem about the existence of solutions of finite difference equations.
The method we use is based on our assumption about the existence of
regular solutions of (\ref{main}) in the neighborhood of the turning
points and they could be approximated by the quasiclassical solutions in
regions far from the turning points.


The remaining singular point is $\rho =0$. Again, we pass to the regular
coordinate on $S_F$ near this point:
 \begin{equation}
 \left\{
 \begin{array}{rcl}
v&=&S^{1/4}\\
P_v&=&4v^3P_s
 \end{array}
 \right.
 \end{equation}
then in terms of this coordinate
 \begin{equation}
F^{1/2}=i\frac{\sqrt{1-v^2}}{v}
 \end{equation}
and the solution of Hamilton-Jacobi equation is
 \begin{equation}
 \label{zero}
P_v=4v^3\ln\left\{ \frac{\displaystyle i}{\displaystyle \sqrt{1-v^2}}
\left( v-\frac{1}{2v}-\frac{M^2}{8m^2
v^3}\pm \frac{M}{2mv}{\cal Z}\right) \right\}.
 \end{equation}
It follows from (\ref{zero}) that $\rho =0$ is a turning point as well as
the horizon because $P_v\rightarrow 0$ with $v\rightarrow 0$ and the
phase $\Omega$ (\ref{faza}) of the wave function
 \begin{equation}
\Omega =v^4(\ln v-1/4)+(regular\ part)
 \end{equation}
remains finite in the vicinity of $\rho =0$. Using the same procedure as
at the horizon we find the behavior of the Stokes lines at this point
using the truncated equation (\ref{ode}). The Hamilton-Jacobi equation for
(\ref{ode}) is
 \begin{equation}
\frac{P_v^2}{16v^6}=\frac{\displaystyle i}{\displaystyle \sqrt{1-v^2}}\left( 
2v-\frac{1}{v}-\frac{M^2}{4m^2
v^3}\right) -2
 \end{equation}
and $v=0$ is the turning point of third order. In accordance with
(\ref{angles}) we draw five Stokes lines from this point. One of the Stokes
lines coincides with $V_-$ ($Arg\ v=\pi /2)$) as far as the argument of
logarithm in (\ref{zero}) becomes zero on this line and the other
four Stokes lines are situated symmetrically with respect to $V_+\bigcup
V_-$ line (see Fig. \ref{stoc1}). Two of them are in $S_+$ component of
$S_F$ and the two others are in $S_-$ component.

The general structure of Stokes lines on the Riemannian surface $S_F$ is
presented on Fig. \ref{gen}. The Stokes lines divide the Riemannian
surface into eight regions. In regions {\bf 2} and {\bf 4} on the picture
the phase of quasiclassical wave function changes in finite limits
$a<\mbox{ Im}(\Omega ) <b$. In  other regions the phase changes so
that $0<|\mbox{ Im}(\Omega )|<\infty$.
Given a Stokes line and a canonical region containing this Stokes line
 we could construct a fundamental system of
solutions in this region.

Now we could construct a quasiclassical  solution of (\ref{main}) in the
whole $S_F$. We will look for the solutions obeying the two following
natural requirements:\\
 \begin{itemize}
 \item
{\bf (A)} the solution must decrease when $\rho\rightarrow\infty$ along the
$R_+$ and $R_-$ lines (we consider the wave function of bound motion
when the shell could not propagate to infinity);
 \item
{\bf (B)} the solution must be an unambiguous function on the Riemannian
surface $S_F$.
 \end{itemize}
 \begin{equation}
 \label{req}
 \end{equation}
So we start from the decreasing solution in the sector {\bf 1} (Fig.
\ref{gen}). The canonical region which contains$R_+$ is $D_1=\mbox{ {\bf 
1} }\bigcup \mbox{ {\bf 2} }\bigcup \mbox{ {\bf 5} }$
with the reference Stokes line $l_1$, then
the decreasing solution is the second solution of the fundamental system:
 \begin{eqnarray}
\Psi =a_1 \Phi^{l_1}_1+b_1\Phi^{l_1}_2,\  \rho\in D_1\nonumber\\  \left(
\begin{array}{c} a_1 \\ b_1 \end{array} \right) =\left( \begin{array}{c} 0
\\ 1 \end{array} \right)
 \end{eqnarray}
In order to find the form of solution on $R_-$ line we should continue the
solution along some path $\gamma$, connecting the points $A$ on $R_+$ and
$B$ on $R_-$ (see Fig. \ref{gen}). To do this we need to cover
$\gamma$ by the set of overlapping canonical regions $D_1$, 
$D_2={\bf 2}\bigcup {\bf 4}\bigcup {\bf
5}\bigcup {\bf 6}$ and $D_3={\bf 3}\bigcup {\bf 2}\bigcup {\bf 5}$ (see
Fig. \ref{gen}) and find the transition matrix between the
fundamental system of solutions $\Phi_1$ and $\Phi_2$ in $D_1$ and
$\tilde\Phi_1,\ \tilde\Phi_2$ in $D_3$:
 \begin{eqnarray}
 T=T_{3,2}\times T_{2,2} \times T_{2,1}\nonumber\\
\left( \begin{array}{c} a_3 \\ b_3
\end{array} \right)=T \left( \begin{array}{c} 0 \\ 1
\end{array} \right)
 \end{eqnarray}
The matrix $T_{2,2}$ arises by the same reason as in
the transition matrix (\ref{trans}). In the canonical region $D_2$ two
different systems of solutions are defined by the Stokes line $l_2$
and different turning points $\rho =\rho_1$ and $\rho =1$.

The solution
$\Phi^{l_5}_2$ is decreasing on $R_-$. So the requirement (\ref{req},A)
takes the form
 \begin{eqnarray}
 \Psi =a_3 \Phi^{l_5}_1+b_3\Phi^{l_5}_2,\ \rho\in D_3\nonumber\\  \left(
\begin{array}{c} a_3 \\ b_3 \end{array} \right) =\left(
\begin{array}{c} 0 \\ 1 \end{array} \right)
 \end{eqnarray}
As far as both turning points involved in the calculation are simple
turning points, the calculations are the same as in the previous section
and we find that in order to obtain the quasiclassical solution decreasing
with $\rho\rightarrow\infty$ along both $R_+$ and $R_-$ lines the
Bohr-Sommerfeld quantization condition must be satisfied on the Stokes
line $l_2$ (Fig. \ref{gen}):
 \begin{equation}
 \label{Bohr}
\frac{1}{\pi \hbar}\int\limits_{l_2} P_sdS=n+\frac{1}{2}
 \end{equation}

Let us note the important property of the constructed solution. Consider
the form of solution on $T_{\pm}$ lines. Let us recall the classical
behavior of the shell in regions $T_\pm$ of space-time. $T_-$ is a
nonstationary region of inevitable contraction and the classical solution
for the shell is just collapsing solutions while the $T_+$ region is the
region of inevitable expansion and the shell trajectory starting from
singularity $\rho=0$ expands up to $\rho=1$ and leaves the $T_+$ region
(Fig. \ref{space}).  The quasiclassical wave function for both
(\ref{main}) and (\ref{ode}) equations expresses through the two basic
solutions
 \begin{equation}
 \label{aaa}
\Psi =\exp\left( \pm\frac{i}{\hbar}\Omega (S)\right)\phi
 \end{equation}
corresponding to in-going and outgoing waves respectively. So it seems
at first sight that at quantum level the shell does not contract
inevitably in $T_-$ region. But
the phase $\Omega (S)=\int P_s dS$ is complex on this line
 \begin{equation}
 \label{ps}
P_s=\ln \left| F^{-1/2}\left( 1+\frac{\displaystyle 1}{ \displaystyle
2\sqrt{S}}+\frac{\displaystyle M^2}{\displaystyle 8m^2S}\pm
\frac{\displaystyle M}{\displaystyle 2m\sqrt{S}}{\cal
Z}\right) \right|-i\mbox{ Arg}(F^{1/2}).
 \end{equation}
In each of the regions
{\bf 1..8} on which the Stokes lines divide the Riemannian surface $S_F$
one of the solutions (\ref{aaa}) is exponentially small compared to the
other.  The function $F^{1/2}$ in the neighborhood of the horizon $\rho =
1$ is given by
 \begin{equation}
F^{1/2}=\frac{u}{\sqrt{1+u^2}}
 \end{equation}
and the argument of $F^{1/2}$ coincides with the argument of the regular
coordinate $u$ on Riemannian surface $S_F$ in the vicinity of $\rho =1$.
The argument of $u$ is zero on the line $R_+$ so the solution is an
oscillating function on this line  near the horizon. On the line $R_-$
the argument of $u$ is $\pm\pi$ so we have decreasing and increasing
components in the fundamental system of solutions in sector {\bf 3} of
Riemannian surface $S_F$. The same situation is on lines $T_\pm$. The
argument of $u$ is $\pm\pi/2$ and the solution $\Psi =\exp (+i\Omega
/\hbar )$ which represents an outgoing wave is decreasing on line $T_-$
while the solution $\Psi =\exp (-i\Omega /\hbar )$ (an in-going wave) is
increasing and therefore is exponentially large with respect to the
outgoing wave on the line $T_-$.  On the line $T_+$ the in-going wave is
exponentially dumped by the same reason. If we require that the solution
is decreasing in sector {\bf 3} of the Riemannian surface $S_F$ and
therefore the in-going and outgoing waves have equal amplitudes on the line
$R_+$,  we conclude that the in-going wave in $T_-$ region and out-going
wave in $T_+$ region both have nonzero amplitudes.
In quasiclassical expression for the wave function we could not
retain the exponentially small part of solution simultaneously with
exponentially large  and therefore could not notice the outgoing wave in
$T_-$ region and the in-going wave in $T_+$ region.


Let us now analyze the requirement (\ref{req},B). The solution of the
equation (\ref{main}) or (\ref{ode}) is defined on the Riemannian surface
$S_F$ (which is topologically two-dimensional sphere) with six punctured
points: $\rho_1, \rho_2, 0, 1$ -- the turning points and two infinities
in $S_{\pm}$ components of $S_F$ which are singular points.
There exist six nontrivial basic cycles $\gamma_1, ...,\gamma_6$ around
each of the punctured points on $S_F$. The procedure
of construction of the quasiclassical solutions in complex domain gives
the wave functions with trivial monodromy along the basic cycles
$\gamma_1,..\gamma_4$ around all the turning points $0,1,\rho_1,\rho_2$.
So we should only require that the wave function transforms into itself
when prolonged along the pass $\gamma_5$ or $\gamma_6$ around the infinities
in $S_\pm$ components of $S_F$. It suffices to
satisfy the condition only on one of the cycles because the other one is
 \begin{equation}
\gamma_6=(\gamma_1\bullet\gamma_2\bullet\gamma_3\bullet\gamma_4\bullet
\gamma_5 )^{-1}
 \end{equation}
(bullet denotes the composition law in fundamental group)
It will be convenient to consider the cycle
$\Gamma$
shown on Fig. \ref{gen} (one could easily check that it is not a 
composition of cycles $\gamma_1,...,\gamma_4$ around the turning points). 
One could see
that we could cover all the pass $\Gamma$ by two canonical regions $D_3={\bf 2}
\bigcup {\bf 5}\bigcup {\bf 3}$ and $D_4={\bf 4}\bigcup {\bf 5}\bigcup {\bf 3}$
(see Fig. \ref{gen}).
Let us prolong the solution from the region {\bf 3}
containing the $R_-$ line and the point $A\in \Gamma$ (Fig.
\ref{gen}) corresponding to the value $t=0$ and $t=1$ of the
parameter along $\Gamma$ to the region {\bf 5} containing $V_+$ line and
the point $B\in\Gamma$ corresponding to $t=1/2$.

The canonical region $D_3$ contains two Stokes lines $l_5$ and $l_8$.
Therefore, there exists two different fundamental systems of solutions in
this region $\Phi^{l_5}_{1,2}$ and $\Phi^{l_8}_{1,2}$. The solution which
decrease in the sector {\bf 3} of $S_F$ is
 \begin{eqnarray}
 \Psi =a_3\Phi^{l_5}_1+b_3\tilde\Phi^{l_5}_2,\ \rho\in D_3\nonumber\\
\left( \begin{array}{c} a_3 \\ b_3 \end{array} \right) =\left(
\begin{array}{c} 0 \\ 1 \end{array}\right)
 \end{eqnarray}
In order to clear up the behavior of this solution in sector {\bf 5} which
belongs to the same canonical region $D_4$ as the sector {\bf 3} we
express the solution $\Psi$ through the fundamental system of solutions
defined by the Stokes line $l_8$ in canonical region $D_4$.  The
transition matrix is
 \begin{equation}
T_{3,3}=\left(\begin{array}{cc} \exp
\left( \frac{i}{\hbar}\int\limits_{1,(T_-)}^0 P_s dS \right) & 0 \\
0 & \exp\left( -\frac{i}{\hbar}\int\limits_{1,(T_-)}^0 P_s dS \right)
\end{array}
\right)
 \end{equation}
Here the integral is taken along the interval $(0,1)$ corresponding to
$T_-$ line on the Riemannian surface $S_F$. So the wave function has the
form
 \begin{equation}
 \label{first}
\Psi =\exp\left( -\frac{i}{\hbar}\int\limits_{1,(T_-)}^0 P_s dS
\right)\Phi^{l_8}_2
 \end{equation}
on the line $V_+$ where $\Phi_2^{l_8}$ is increasing when $\rho\rightarrow
-\infty$ along $V_+$.

Let us consider the form of wave function in the region $D_4$. There are
two Stokes lines $l_4$ and $l_9$ in this region.
In terms of fundamental system of solutions defined by the
Stokes line $l_5$ the wave function has the form:
 \begin{eqnarray}
 \Psi =a_4\Phi^{l_5}_1+b_4\Phi^{l_5}_2,\ \rho\in D_4\nonumber\\ \left(
\begin{array}{c} a_4 \\ b_4 \end{array} \right) =\left(
\begin{array}{c} 1 \\ 0 \end{array}\right)
 \end{eqnarray}
When expressed through the fundamental system of solutions defined
by the Stokes line $l_9$ the wave function becomes
 \begin{equation}
 \label{second}
\Psi =\exp\left( \frac{i}{\hbar}\int\limits_{1,(T_+)}^0 P_s dS \right)\Phi^{l_9}_1
 \end{equation}
where $\Phi^{l_9}_1$ has the property $\Phi^{l_9}_1(\rho )\rightarrow
+\infty$ when $\rho\rightarrow -\infty$ in $V_+$ region.

From the symmetry of the construction with
respect to real line we have
 \begin{equation}
\Phi^{l_8}_2 (\rho)=\Phi^{l_9}_1 (\rho),\ \rho\in V_+
 \end{equation}
so the two expressions for the wave function (\ref{first}) and
(\ref{second}) agree if
 \begin{equation}
\exp\left(
\frac{i}{\hbar}\left(\int\limits^0_{1,T_-}+\int\limits^0_{1,T_+}\right)
P_s dS \right) =1
 \end{equation}
Taking in account the fact that the wave function is in-going wave on $T_-$
line and outgoing on $T_+$ line, as it is explained earlier, we see that
the integral in the argument of the exponent could be written as the
integral over the closed path which goes around the cut made from the
point $\rho =0$ to $\rho =1$ on both of the Riemannian spheres $S_\pm$
constituting the Riemannian surface $S_F$.  So the quasiclassical wave
function satisfy the requirement (\ref{req},B) if the quantization
condition
 \begin{equation}
 \label{Sommerfeld}
\frac{1}{2\pi\hbar}\left(\ointop^1_0 P_s dS \right)=k,\ k\in Z
 \end{equation}
holds.

Collecting the results obtained in this section we conclude that in zero
order on $\hbar$ the quasiclassical solution of equation (\ref{main})
defined on the Riemannian surface $S_F$ exists if the two quantization
conditions (\ref{Bohr}) and (\ref{Sommerfeld}) hold. Taking in account
the explicit expression (\ref{solution}) for $P_s$ and calculating the
integrals entering the quantization conditions we obtain the equations
defining the quasiclassical spectrum for the bound motion of
the selfgravitating dust shell:
\widetext
 \begin{eqnarray}
\int\limits_1^{\rho_1^2} P_s dS=-\frac{ \displaystyle 1}{\displaystyle 2}
+\frac{ \displaystyle M^2}{\displaystyle 8 m^2} 
+\frac{\displaystyle
3}{\displaystyle 2}\ln\left(\frac{ \displaystyle 2 m}{\displaystyle
M}\right) 
-\frac{\displaystyle
2-\frac{M^2}{m^2}}{\displaystyle 
4\sqrt{ \frac{M^2}{m^2}-1}} \left( \pi +3\mbox{arccos} \left(
\frac{\displaystyle m}{\displaystyle M} \right) \right)\nonumber \\
\int\limits_0^{1} P_s dS=\frac{\displaystyle 1}{\displaystyle 2}-
\frac{
\displaystyle M^2}{\displaystyle 4 m^2}-
\ln\left( \frac{\displaystyle 2
m}{\displaystyle M}\right) +
\frac{\displaystyle 
2-\frac{M^2}{m^2}}{\displaystyle 2 \sqrt{
\frac{M^2}{m^2}-1}}\mbox{arccos}
\left(\frac{\displaystyle m}{\displaystyle M}\right)
 \end{eqnarray}
Denoting $\cos \alpha =m/M$ we obtain
 \begin{equation}
 \label{BZ}
 \left\{
 \begin{array}{lc}
f_1=\frac{\displaystyle 1}{\displaystyle 2}-
\frac{\displaystyle 1}{\displaystyle 4 \cos^2\alpha}-
\ln\left( 2 \cos\alpha \right) +
\alpha\mbox{ ctg } 2\alpha
=\pi\zeta k & k\in Z\\
f_2=-\frac{\displaystyle 1}{\displaystyle 2}
+\frac{\displaystyle 1}{\displaystyle 8 \cos^2\alpha}+
\frac{\displaystyle 3}{\displaystyle 2}\ln\left( 2 \cos\alpha \right) -
\frac{\displaystyle \left( 3\alpha +\pi\right) \mbox{ ctg }
2\alpha}{\displaystyle 2} =\pi\zeta \left( n+
\frac{ \displaystyle 1}{\displaystyle 2}\right) & n\in Z\\
 \end{array}
 \right.
 \end{equation}
\narrowtext


The behavior of the functions standing in the left hand side of this
equation is shown on Fig. \ref{graph}. Large $k$ and large positive $n$
correspond to wormhole states while large negative $n$ and finite $k$
correspond to the black hole states as one could see.

Let us consider the spectrum in the black hole case. This corresponds
to the values $m<M<2m$ of parameters on Fig. \ref{graph}a. In this case
the function $f_1$ changes in finite limits from
$f_1=\pi/(3\sqrt{3})-1/2$  at $m/M=1/2$ to $f_1=3/4-\ln 2$ at
$m/M=1$.  So if the Schwarzschild mass $m$ is given, then $k$ could change
only in finite limits
 \begin{equation}
 \label{finite}
\frac{\displaystyle 1}{\displaystyle \zeta}\left(
\frac{1}{3\sqrt{3}}-\frac{1}{2\pi}\right)
<k<\frac{1}{\pi\zeta}\left(\frac{3}{4}-\ln 2\right)
 \end{equation}
So the black hole state with given Schwarzschild mass
$m$ (the only parameter that an observer at infinity could measure) is a
superposition of the states with different $k$ and is in fact a mixed
state, having nonzero entropy. Besides, we have the inequality
 \begin{equation}
m>m_{pl}/\sqrt{ \left(
\frac{1}{6\sqrt{3}}-\frac{1}{4\pi}\right)}
 \end{equation}
for the quasiclassical black hole spectrum because otherwise the
inequality (\ref{finite}) have no solutions. This means that the minimal
black hole mass exists (if we suppose that quasiclassical spectrum is valid for
low energy values).

The mass spectrum for the black hole model under consideration was found
in \cite{prd57} in the large black hole limit using another technic. Let
us compare the two spectra. From  (\ref{BZ}) we find
 \begin{equation}
 \label{combination}
\frac{\displaystyle \frac{\displaystyle M^2}{\displaystyle m^2}-1}{
\displaystyle 2\pi\zeta}+\frac{\displaystyle 2-\frac{M^2}{m^2}}{
\displaystyle 2\zeta\sqrt{\frac{M^2}{m^2}-1}}=-(3k+2n+1)
 \end{equation}
We see that although each of quantization conditions found in
\cite{prd57} does not hold for the quasiclassical spectrum, their
combination (\ref{combination}) does hold.

Let us consider the limit $M>m$, $m\rightarrow M$ which corresponds to the
case when the turning point of the classical motion of the shell
$\rho_1\rightarrow\infty$. In this limit $\alpha\rightarrow 0$ and the
last item in the second of equations (\ref{BZ}) becomes large compared to
the others. Then the second quantization condition (\ref{BZ}) takes the
form in this limit
 \begin{equation}
\frac{\displaystyle 2-\frac{M^2}{m^2}}{
\displaystyle 4\zeta\sqrt{\frac{M^2}{m^2}-1}}=n+\frac{\displaystyle 1}{
\displaystyle 2}
 \end{equation}
which coincides qualitatively with the quantization condition of
\cite{prd57}.  So the two spectra coincide in the limit $m\rightarrow M$.

\section{The quasiclassical spectrum for the collapsing shell.}
\label{sec:collapse}

In the previous section we obtained the discrete mass spectrum for the
bound motion of the selfgravitating shell. One of the quantization
conditions (\ref{req},A) appeared to be similar to the
Bohr-Sommerfeld quantization condition on the trajectories of the bound
motion of mechanical system in nonrelativistic quantum mechanics.
The other quantization condition (\ref{req},B) has a different
origin being the requirement for the wave function to be a regular
function on the Riemannian surface $S_F$. The first requirement
(\ref{req},A) will disappear if we consider the situation of
gravitational collapse when the classical trajectory goes from
$\rho =\infty$ to $\rho =0$ (see Fig.\ref{space}c). The
wave function is the superposition of in-going and out-going waves near
$\infty$ and does not decrease in this region.  So the first of the
quantization conditions (\ref{Bohr}), (\ref{Sommerfeld}) disappear. The
principally new feature of our model compared to the ordinary
nonrelativistic quantum mechanics is that the second condition
(\ref{Sommerfeld}) which stems from the requirement (\ref{req},B) for
the wave function to be an unambiguous function on the Riemannian
surface $S_F$
does not disappear in the case of unbound motion and gives
the discrete spectrum for the mass of the collapsing shell.

The collapse case corresponds to the values of parameters $m>M$. In this
case the integral entering the quantization condition (\ref{Sommerfeld})
takes the form
 \begin{equation}
\int\limits_0^{1} \mbox{ Re }(P_s) dS=\frac{\displaystyle 1}{\displaystyle
2}-\frac{ \displaystyle M^2}{\displaystyle 4 m^2}-\ln\left(
\frac{\displaystyle 2 m}{\displaystyle M}\right) +\frac{\displaystyle
2-\frac{M^2}{m^2}}{ \displaystyle 2\sqrt{1-\frac{M^2}{m^2}}}\mbox{ arcch}
\left(\frac{\displaystyle m}{\displaystyle M}\right)
 \end{equation}
Denoting
$\mbox{ ch}(\alpha )=m/M$ we obtain
 \begin{equation}
 \label{BZcoll}
\tilde f_1=\frac{\displaystyle 1}{\displaystyle
2}-\frac{ \displaystyle 1}{\displaystyle 4 \mbox{ ch}^2(\alpha )}-
\ln\left( 2\mbox{ ch}(\alpha )\right)
+\alpha \mbox{ cth}(2\alpha )=\pi\zeta k,\ k\in Z
 \end{equation}

The graph of this function is presented on Fig. \ref{graph}b. We see that
when the Schwarzchild mass is fixed there are only finite number of
quantum states of the collapsing shell, because $\tilde f_1$ changes from
$3/4-\ln 2$ to $1/2$ for all the values $1<m/M<\infty$. So
 \begin{equation}
 \label{finite1}
\frac{\displaystyle 1}{\displaystyle \pi\zeta}\left(\frac{3}{4}-\ln
2\right) <k<\frac{1}{2}
 \end{equation}
The
spectrum is continuous as far as it depends on a continuous parameter
$M$ but each level $m=const$ is finite degenerate similarly to the
black hole case. Similarly to the black hole case we see that equation
(\ref{finite1}) has  solutions only if
 \begin{equation}
 \label{minimal}
m> \sqrt{\pi}m_{pl}
 \end{equation}
which means that the minimal mass exists for the states of the collapsing
shell in the quasiclassical spectrum.

Let us consider now the collapsing null-dust shell ( $M=0$ ).
We write according to the Dirac quantization procedure the quantum version
of the constraint (\ref{iso}). It is the following finite difference
equation:
 \begin{equation}
\Psi (S+i\zeta)+\Psi (S-i\zeta) =F^{-1/2}\left( 2-
\frac{\displaystyle 1}{\displaystyle\sqrt{S}}\right)
 \end{equation}
We look for the quasiclassical solutions on the whole $S_F$ of this
equation. The corresponding Hamilton-Jacobi equation
 \begin{equation}
\mbox{ ch }P_s=F^{-1/2}\left( 1-\frac{\displaystyle 1}{\displaystyle
2\sqrt{S}}\right)
 \end{equation}
has the solutions
 \begin{equation}
P_s =\pm\ln\left( F^{1/2}\right)
 \end{equation}
The integral entering the quantization condition (\ref{Sommerfeld}) is
 \begin{equation}
\int\limits_0^1 \mbox{ Re }(P_s) dS=\pm\frac{\displaystyle
1}{\displaystyle 2}
 \end{equation}
and (\ref{Sommerfeld}) gives the spectrum
 \begin{equation}
m^2=\pi m_{pl}^2 k, \ k\in Z
 \end{equation}
proposed by Bekenstein, Mukhanov and other authors.

Let us consider the quasiclassical wave function
in the case $m>M$. We would expect that the quasiclassical wave function
will be concentrated near the classical trajectory of the collapsing
shell in the phase space. Indeed, the in-going wave quasiclassical
solution corresponds to this trajectory. But it turns out that we could
not set the amplitude of the out-going wave to be equal to zero. We
require that the wave function in $R_-$ region to be equal to
the decreasing solution of (\ref{main}) (otherwise it grows infinitely
with $\rho\rightarrow\infty$ in $R_-$ region).  Then we
prolong of the solution to the whole Riemannian
surface $S_F$ following the procedure described in sections
\ref{sec:fedor} and \ref{sec:bh}. We find that the in- going and out-going
waves enter the wave function with equal amplitudes on $R_+$ line.
Therefore we see that the situation resembles the reflection from a
potential wall in nonrelativistic quantum mechanics. The horizon point
$\rho =1$ is a turning point $P_u=0$ for the quasiclassical motion if we
choose the regular coordinate $u$ on the Riemannian surface $S_F$ near the
horizon. So the equation (\ref{main}) describes the situation when
the stationary in-going wave reflects completely from the horizon. In
$T_-$ region the wave function is an in-going wave and in $T_+$ it is an
out-going wave similarly to the case of bound motion.

 \section{Conclusion.}
 \label{sec:conclusion}

In this paper we considered the quantum mechanical model of the black hole
that consists of the selfgravitating thin dust shell. The Schroedinger
equation for this model is a finite difference equation (\ref{main})
with
the finite shift of the argument of the wave function along the imaginary
axis \cite{prd57}.  Therefore, the equation must be considered on a
Riemannian surface $S_F$ where the function $F^{1/2}$ (\ref{F}) is regular.
The analysis of the equation on this Riemannian surface gives the mass
spectrum for the black hole model in the quasiclassical approximation.
While constructing the  quasiclassical solutions on a complex manifold
we need to determine the approximate solutions of the differential
equation in different canonical domains of the complex manifold. Then we
glue the global solution from the solutions defined in different
canonical regions with the help of transitions matrices between these
regions. In the section \ref{sec:bh} this method was used in order to
find the quasiclassical solutions of the equation (\ref{main}). The
quasiclassical solution satisfying the two requirements (\ref{req},A,B)
exists if the coefficients of the equation satisfy two quantization
conditions (\ref{Bohr}) and (\ref{Sommerfeld}).  One of the obtained
quantization conditions (\ref{BZ}) follows from the boundary conditions at
infinities in $R_\pm$ regions of Kruskal space-time ( requirement
 (\ref{req},A))
and is obtained by
the same procedure as Bohr-Sommerfeld condition of nonrelativistic quantum mechanics.
But the other (\ref{BZ}) which follows from (\ref{req},B) and is expressed
as the requirement of the trivial monodromy of the wave function along
nontrivial cycle $\Gamma$ (Fig.  \ref{gen}) on the Riemannian surface
$S_F$ with punctured singular points of the equation (\ref{main}). It has
no analogy in ordinary quantum mechanics.

Being combined together the quantization conditions (\ref{BZ}) define the
discrete mass spectrum of the model, which depends on two quantum numbers.
Furthermore, one of the quantization conditions (\ref{BZ}) remains valid
for the unbound motion where it takes the form (\ref{BZcoll}).
The spectrum of unbound motion
is continuous but each level is finite degenerate. The black hole spectrum of
both bound and unbound motions are bounded from below $m>\sqrt{\pi}m_{pl}$.
For the null-dust $M=0$ the remaining quantization condition gives the
discrete spectrum of unbound motion which turns out to be the spectrum
found by Bekenstein and Mukhanov \cite{bekenstein}.

\acknowledgements

I wish to thank V.A.Berezin, A.M.Boyarsky and A.I.Shafarevich for many
fruitful discussions of the question considered in this paper.
This work was supported in part by the Russian Fund
for Fundamental Research (Grant No 97-02-17064-a) and by Institute "Open
Society" (Grant No a98-1857).

 \begin{figure}
 \caption{The space-time with selfgravitating thin shell. The
          black hole case (a) the wormhole case (b) and the case of
          collapsing shell (c).}
 \label{space}
 \end{figure}

 \begin{figure}
 \caption{The structure of Stokes lines for the potential with
          two turning points $z_0$ and $z_1$.}
 \label{schro}
 \end{figure}

 \begin{figure}
 \caption{The real section of the Riemannian
surface $S_F$ is a cross naturally identified with the
configuration space of classical dynamical system.}
 \label{penrose}
 \end{figure}

 \begin{figure}
 \caption{The Stokes lines on $S_F$ near the turning points
          $\rho_1$ and $\rho_2$.}
 \label{stoc2}
 \end{figure}

 \begin{figure}
 \caption{The Stokes lines near the singular points
          $\rho =\infty$ in $S_\pm$.}
 \label{stoc3}
 \end{figure}

 \begin{figure}
 \caption{The Stokes lines in the vicinity of the horizon
          $\rho =1$ and near the singularity $\rho =0$.}
 \label{stoc1}
 \end{figure}

 \begin{figure}
 \caption{The global structure of Stokes lines on the Riemannian
          surface $S_F$.}
 \label{gen}
 \end{figure}

 \begin{figure}
 \caption{Functions defining the mass
spectrum.}
 \label{graph}
 \end{figure}

 
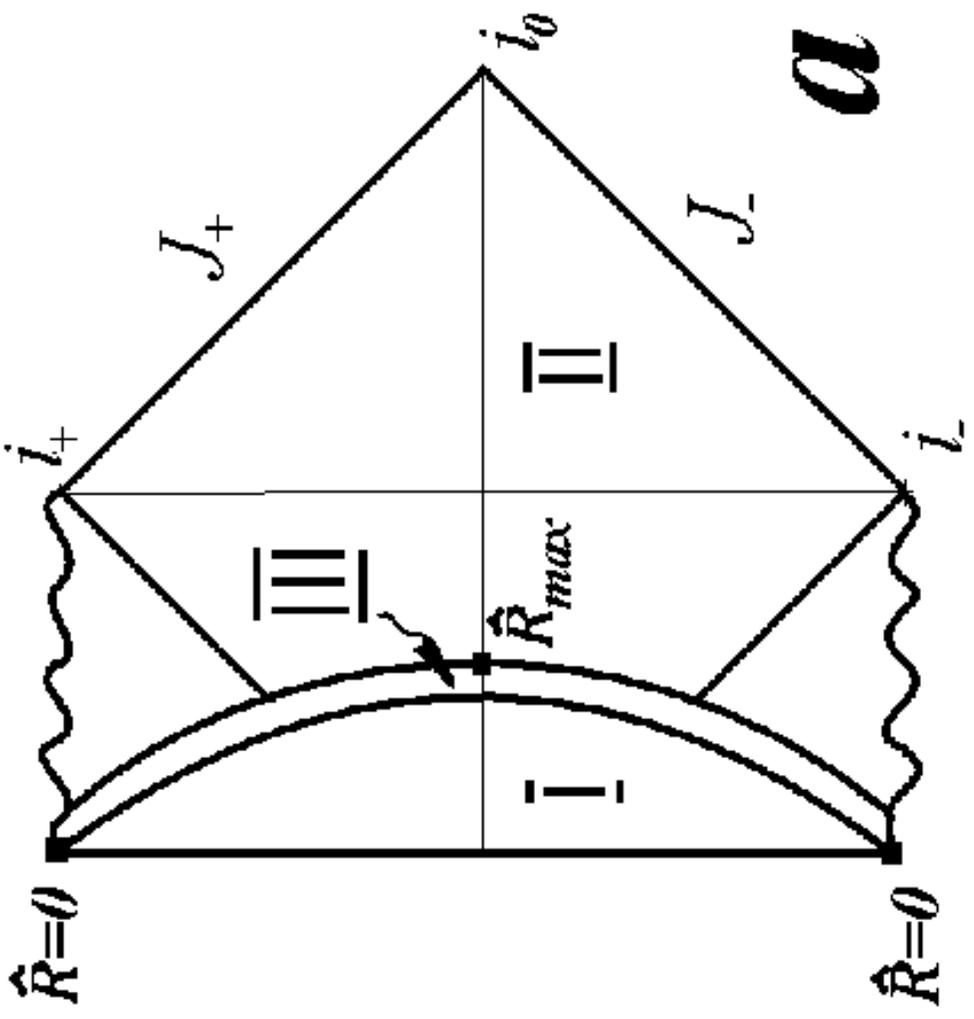
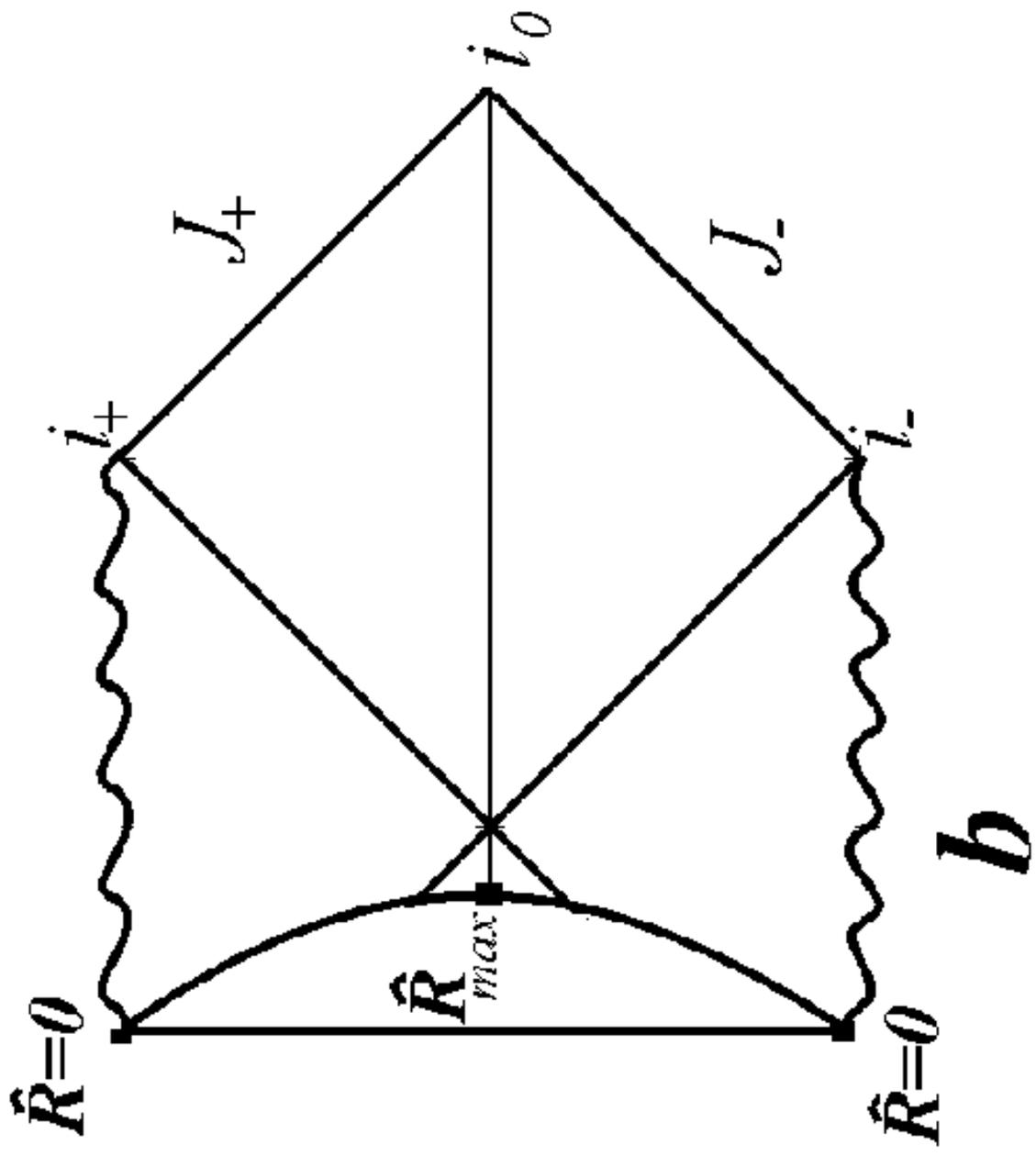
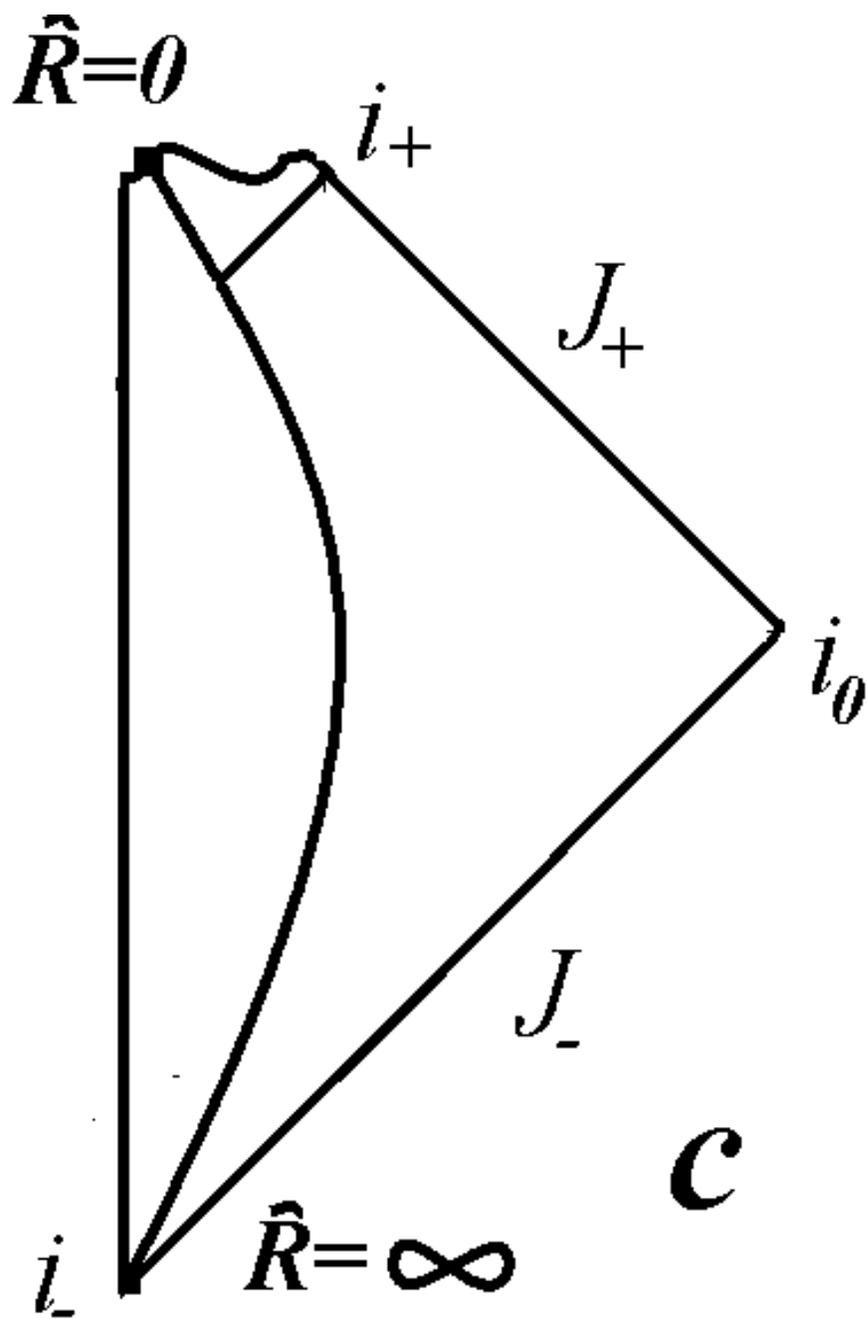
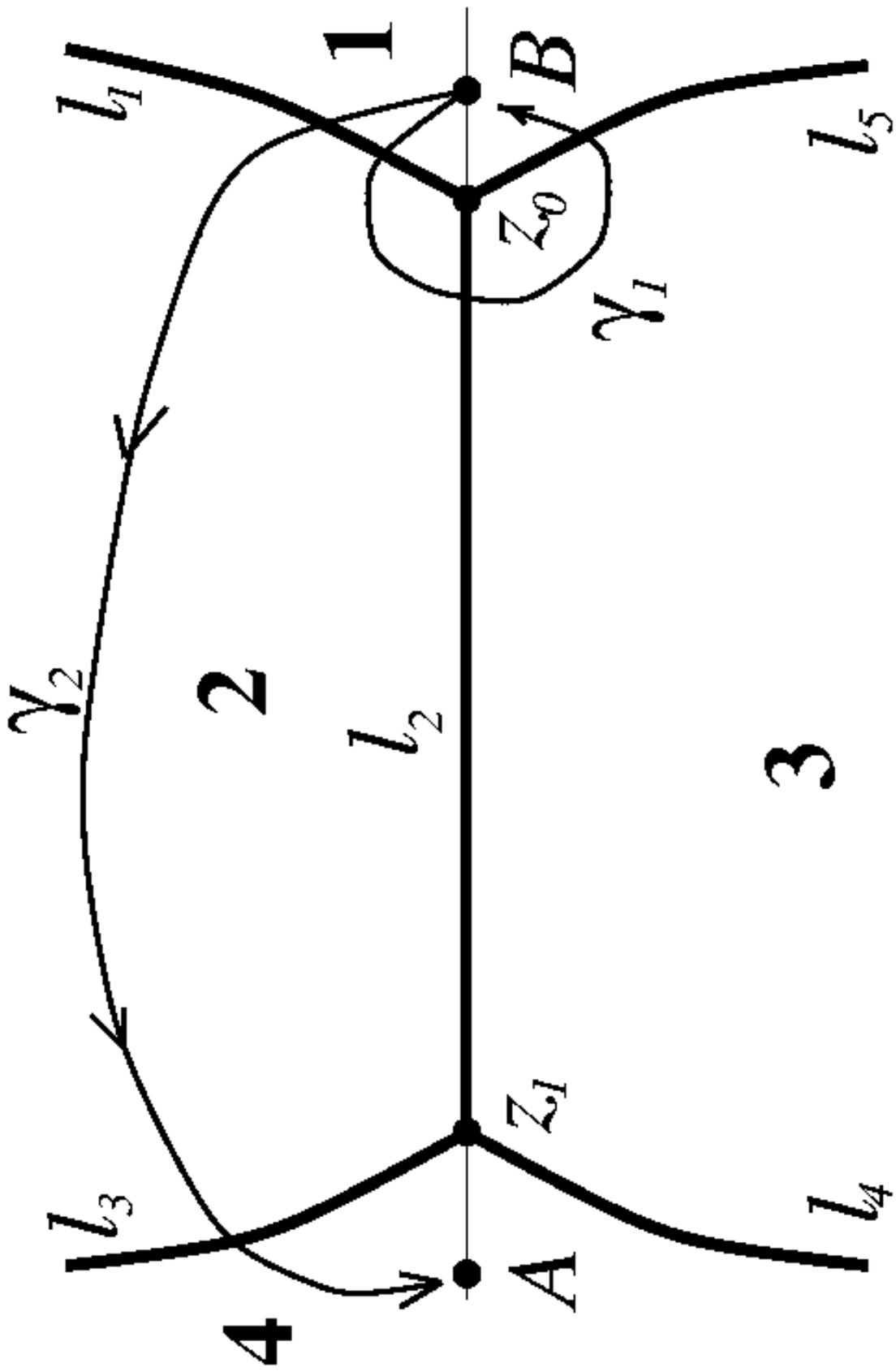
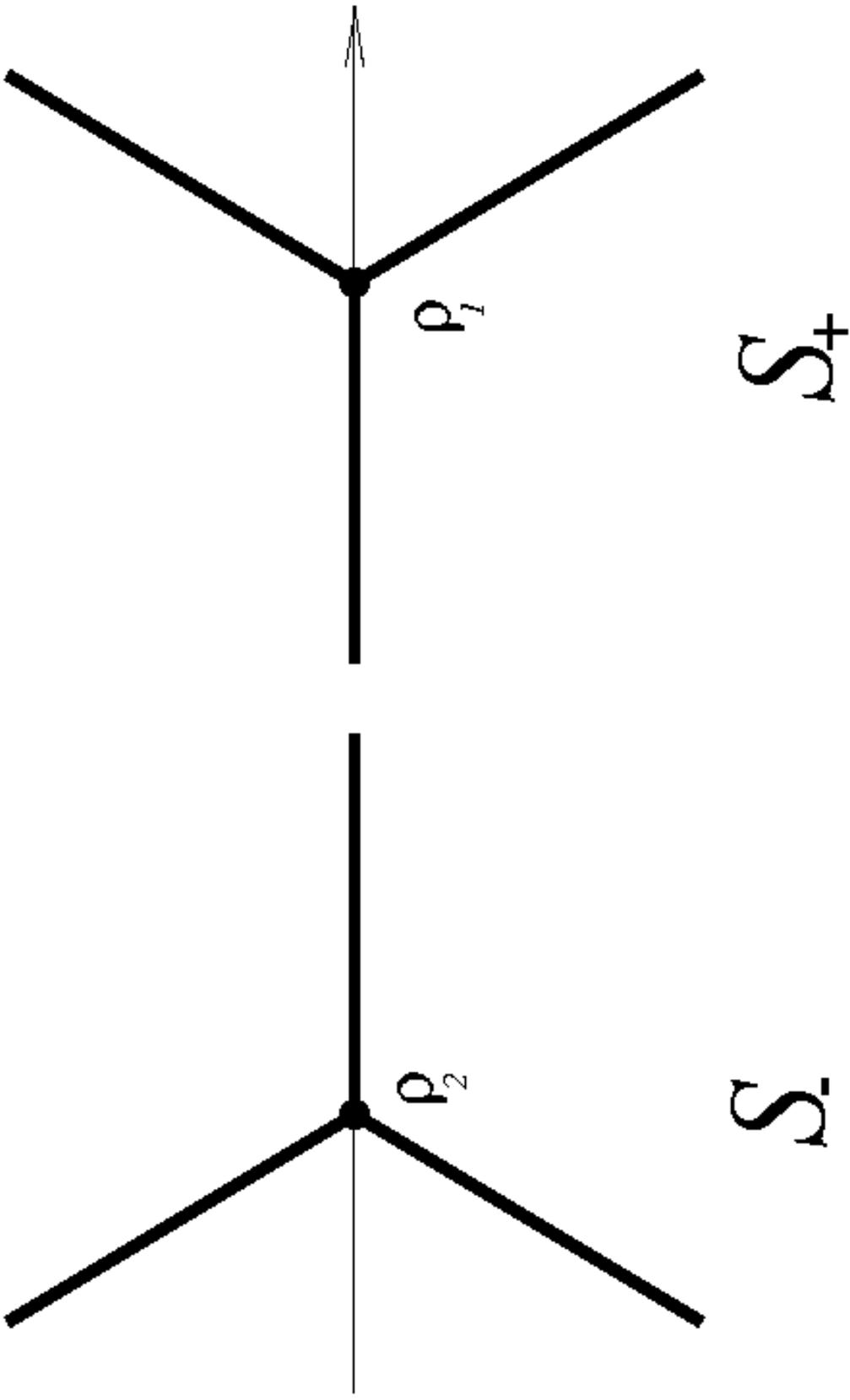
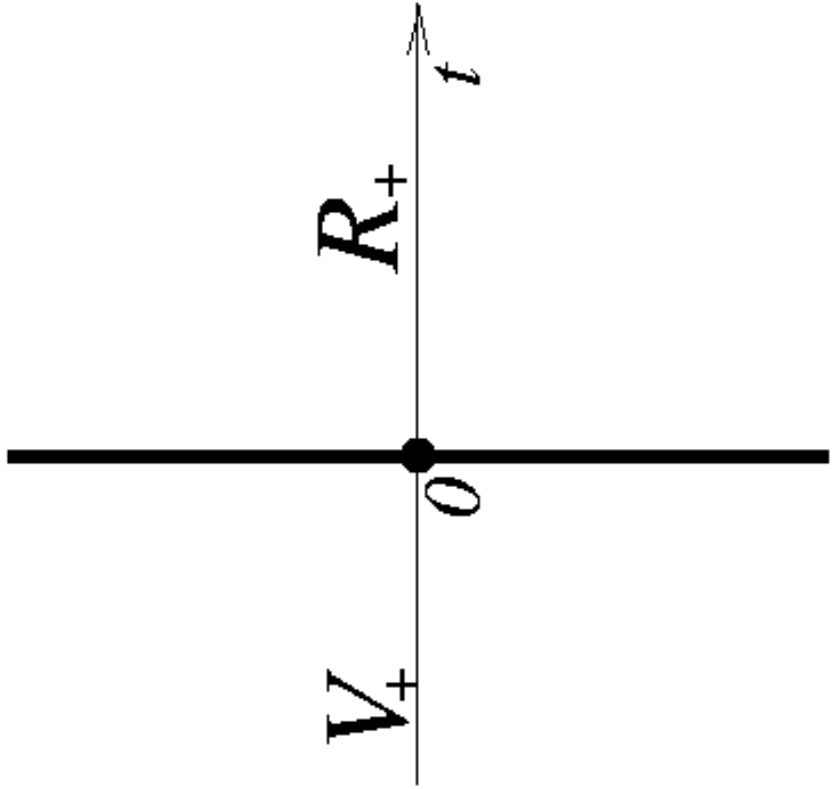
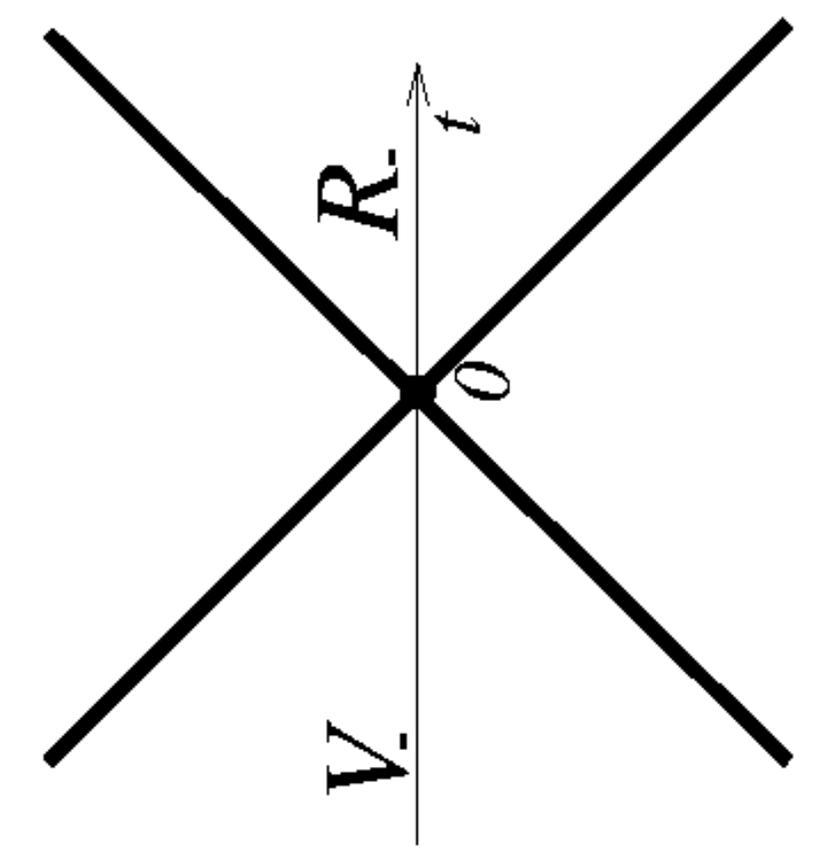
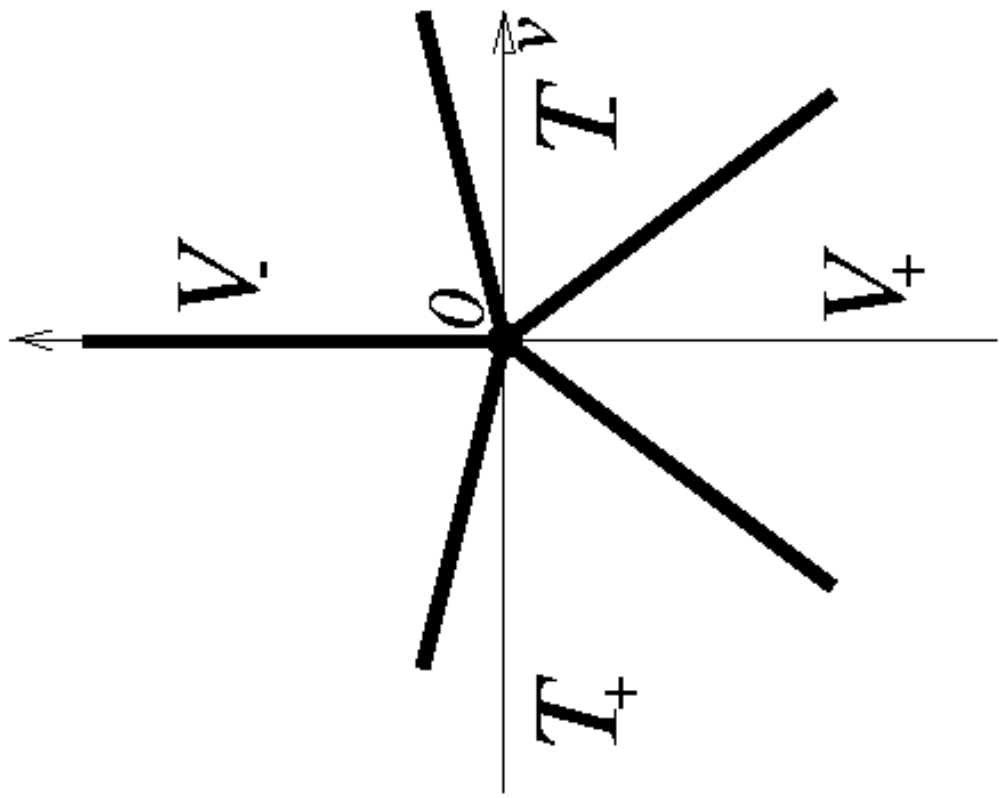
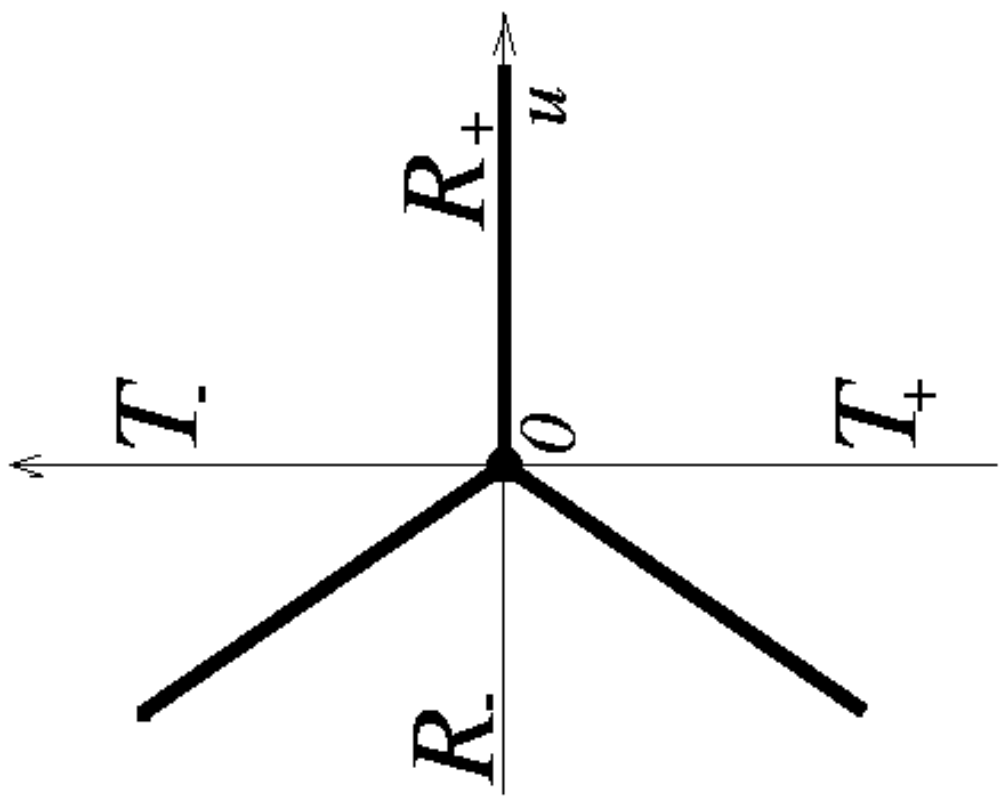
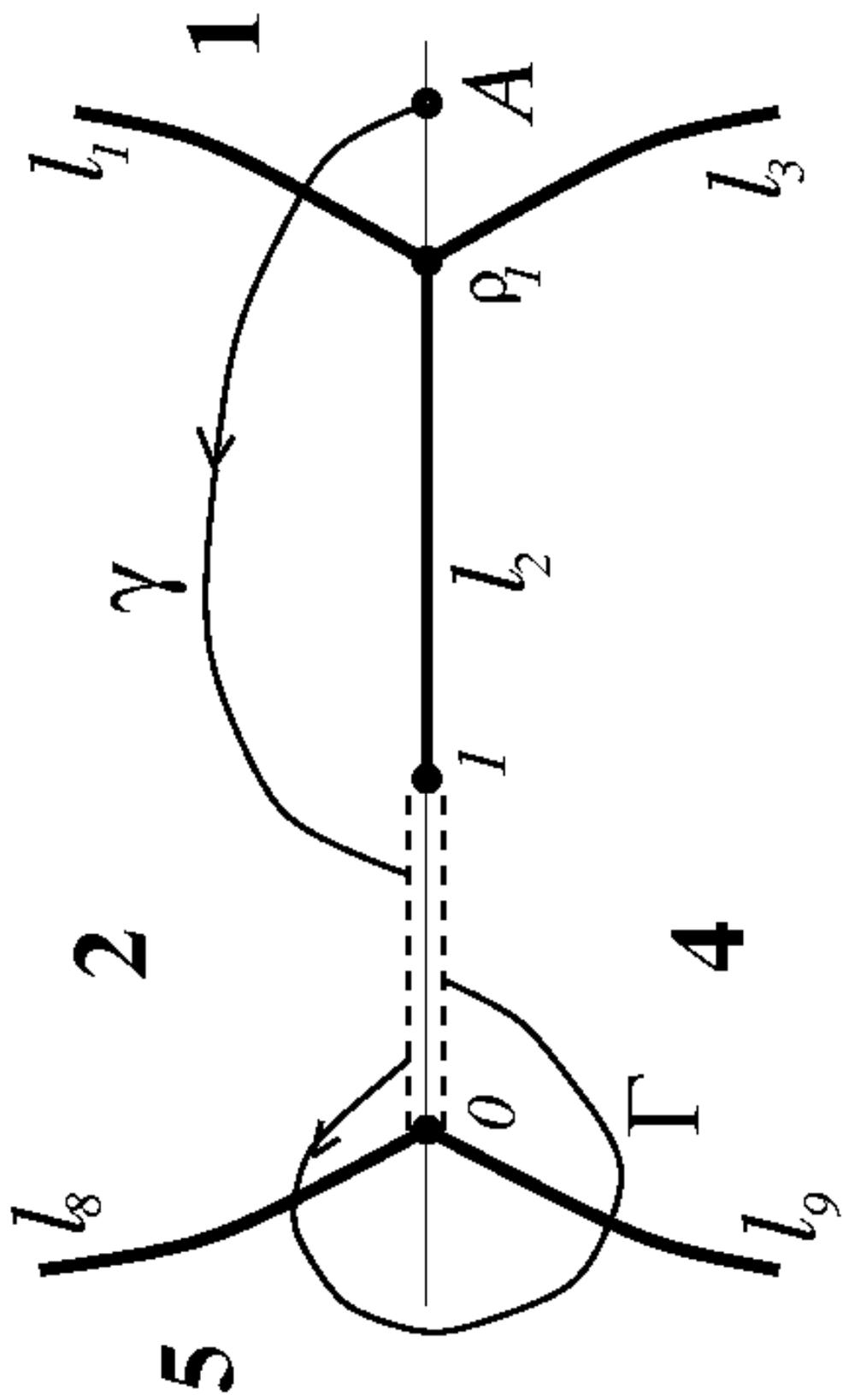
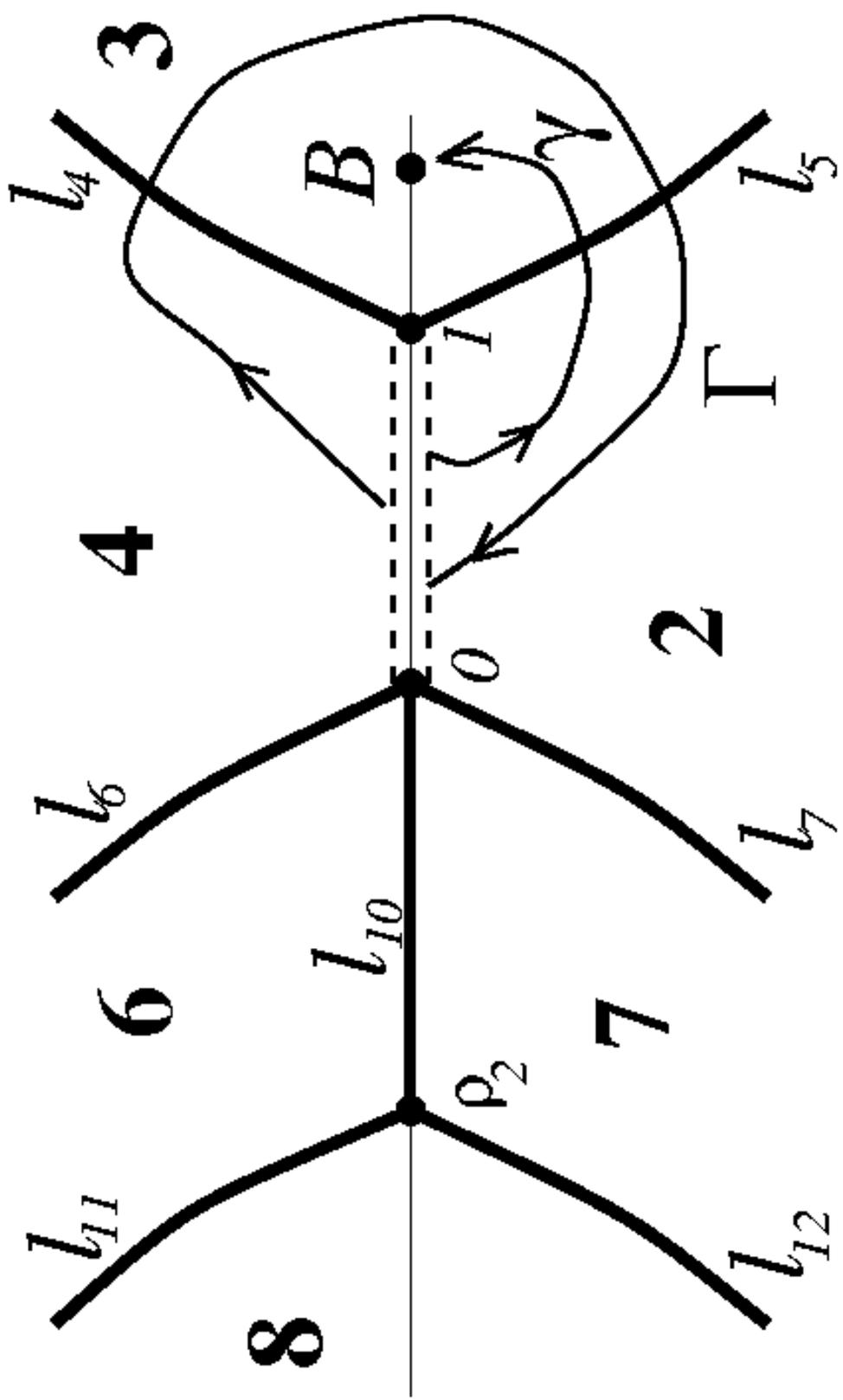
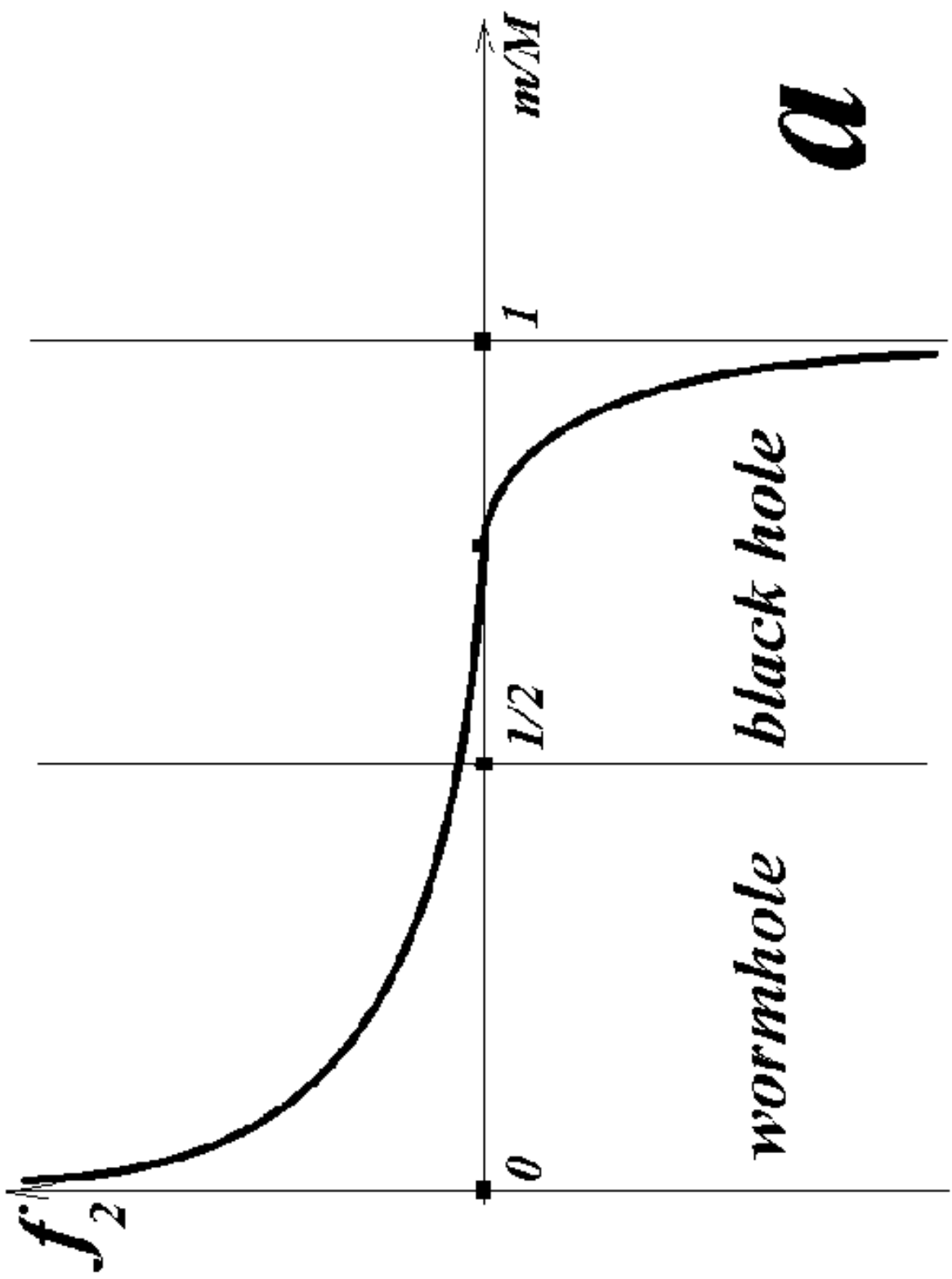
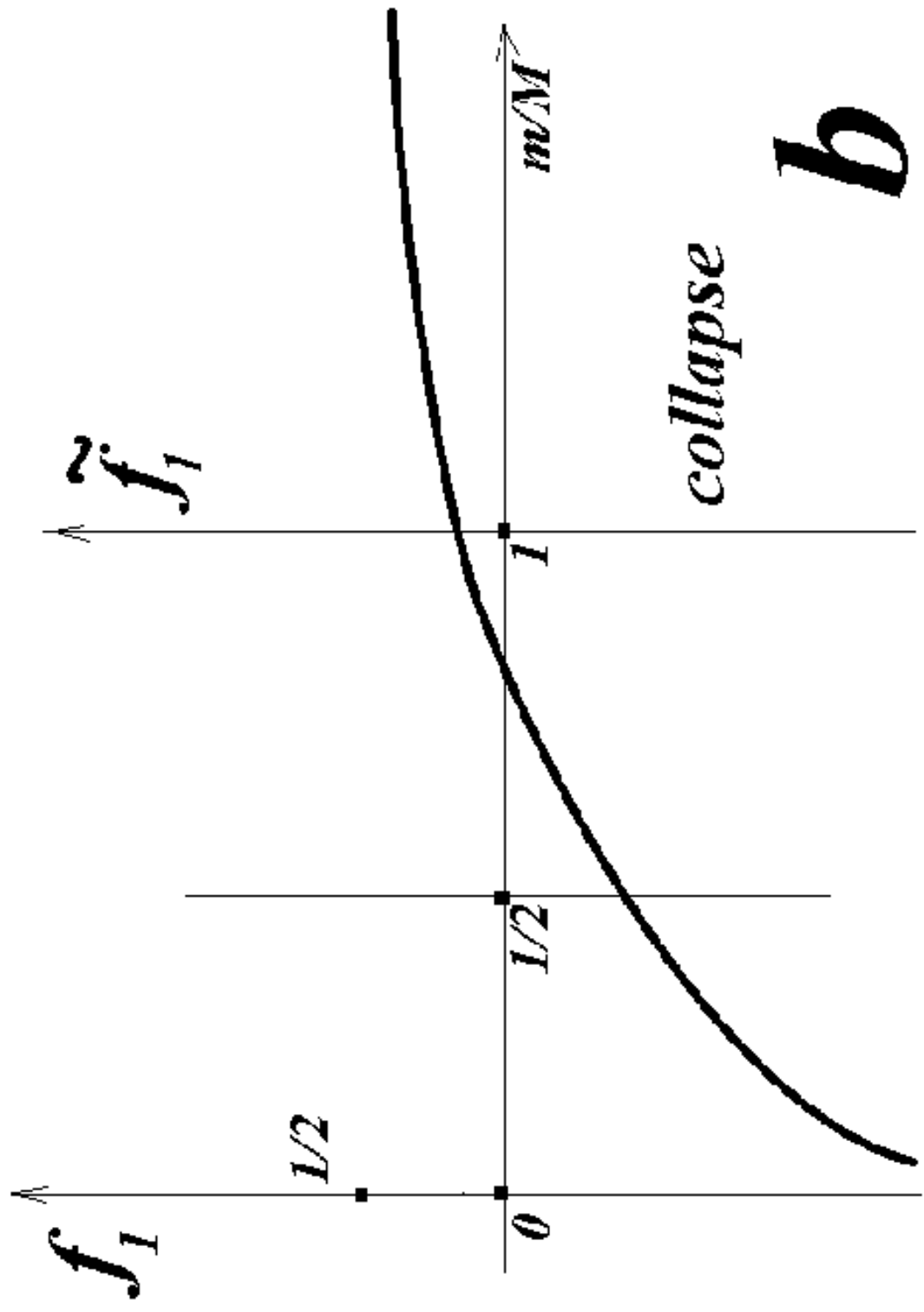
\end{document}